\documentclass[longauth]{aa}
\usepackage[varg]{txfonts}
\usepackage{graphicx}

\usepackage{natbib}
\bibpunct{(}{)}{;}{a}{}{,}
\usepackage[colorlinks,linkcolor=blue,urlcolor=blue,citecolor=blue]{hyperref} 
\usepackage{threeparttablex}
\usepackage{longtable}
\usepackage{lscape}
\usepackage{bm}
\usepackage{verbatim}
\usepackage{makecell}
\usepackage{caption}
\usepackage{subcaption}
\usepackage{multirow}
\usepackage{booktabs}
\usepackage{hyperref}
\usepackage{threeparttable}
\usepackage{colortbl}
\usepackage{xcolor}
\usepackage{booktabs}
\usepackage{float}

\usepackage{txfonts}

\usepackage{xcolor}
\definecolor{myred}{RGB}{255,0,0}
\definecolor{myblue}{RGB}{51,153,255}
\definecolor{myteal}{RGB}{0,179,179}
\definecolor{mypurple}{RGB}{153,51,255}

\usepackage{tikz}
\usetikzlibrary{shapes, positioning, patterns}

\begin{document}

\title{Cosmic Evolution Survey: LOFAR 144 MHz imaging of the COSMOS field}
\titlerunning{LOFAR 144 MHz imaging of COSMOS}

\author{
E. Vardoulaki\inst{\ref{noa},\ref{tls}}\texorpdfstring{\href{https://orcid.org/0000-0002-4437-1773}{\protect\includegraphics{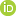}}}{0000-0002-4437-1773},
T. Shimwell\inst{\ref{astron},\ref{leiden}}, 
M. J. Hardcastle\inst{\ref{hertz}},
L. Doukas\inst{\ref{thessaly}}\texorpdfstring{\href{https://orcid.org/0009-0003-3640-1551}{\protect\includegraphics{ORCID-iD_icon-16x16.png}}}{0009-0003-3640-1551},
L. K. Morabito\inst{\ref{durham},\ref{durham_cosm}},
C. L. Hale\inst{\ref{Edinburgh}},\texorpdfstring{\href{https://orcid.org/0000-0001-6279-4772}{\protect\includegraphics{ORCID-iD_icon-16x16.png}}}{0000-0001-6279-4772},
M. T. Sargent\inst{\ref{mark}}\texorpdfstring{\href{https://orcid.org/0000-0003-1033-9684}{\protect\includegraphics{ORCID-iD_icon-16x16.png}}}{0000-0003-1033-9684},
V. Smol{\v{c}}i{\'c}\inst{\ref{zagreb}},
G. Zamorani\inst{\ref{INAF}}, 
G. Gozaliasl\inst{\ref{Aalto},\ref{Helsinki}}, 
F. An\inst{\ref{china}}, 
M. Aravena\inst{\ref{Arav1},\ref{Arav2}}, 
A. L. Faisst\inst{\ref{Caltech}}\texorpdfstring{\href{https://orcid.org/0000-0002-9382-9832}{\protect\includegraphics{ORCID-iD_icon-16x16.png}}}{0000-0002-9382-9832}, 
M. Vaccari\inst{\ref{vac1},\ref{vac2},\ref{vac3}}\texorpdfstring{\href{https://orcid.org/0000-0002-6748-0577}{\protect\includegraphics{ORCID-iD_icon-16x16.png}}}{0000-0002-6748-0577}, 
E. Treister\inst{\ref{Chile}},
A. Finoguenov\inst{\ref{Helsinki}}, 
P. Best\inst{\ref{Edinburgh}}, 
J. S. Kartaltepe\inst{\ref{rochester}}, 
C. Casey\inst{\ref{SB}}, 
S. Jin\inst{\ref{spatiale}}, 
and V. Mainieri\inst{\ref{ESO}}
}

\authorrunning{Vardoulaki et al.}

\institute{
IAASARS/National Observatory Athens, Hill of the Nymphs, Athens, 11851, Greece \label{noa}
\and
Thüringer Landessternwarte Tautenburg, Sternwarte 5, 07778 Tautenburg, Germany \label{tls}
\and
ASTRON, the Netherlands Institute for Radio Astronomy, Postbus 2,
NL-7990 AA Dwingeloo, The Netherlands \label{astron}
\and
Leiden Observatory, Leiden University, PO Box 9513, 2300 RA Leiden,
The Netherlands \label{leiden}
\and
Centre for Astrophysics Research, University of Hertfordshire, Hatfield, Herts, AL10 9AB, UK \label{hertz}
\and
University of Thessaly, 382 21 Volos, Greece \label{thessaly}
\and
Centre for Extragalactic Astronomy, Department of Physics, Durham University, South Road, Durham DH1 3LE, UK \label{durham}
\and
Institute for Computational Cosmology, Department of Physics, Durham University, South Road, Durham DH1 3LE, UK \label{durham_cosm}
\and
Institute for Astronomy, University of Edinburgh, Royal Observatory, Blackford Hill, Edinburgh EH9 3HJ, UK \label{Edinburgh}
\and
Laboratoire d’Astrophysique (LASTRO), EPFL SB IPHYS, Route de la Sorge (Cubotron building), CH-1015 Lausanne, Switzerland \label{mark}
\and
Department of Physics, University of Zagreb, Bijenička cesta 32, Zagreb, Croatia \label{zagreb}
\and
INAF -- Osservatorio di Astrofisica e Scienza dello Spazio di Bologna, Via Piero Gobetti 93/3, I-40129 Bologna, Italy \label{INAF}
\and
Department of Computer Science, Aalto University, PO Box 15400, 00076, Espoo, Finland \label{Aalto}
\and
Department of Physics, University of Helsinki, Gustaf H\"{a}llstr\"{o}m katu 2, 00560 Helsinki, Finland \label{Helsinki}
\and
Yunnan Observatories, Chinese Academy of Sciences, Kunming 650216, People’s Republic of China \label{china}
\and
Instituto de Estudios Astrof\'{\i}cos, Facultad de Ingenier\'{\i}a y Ciencias, Universidad Diego Portales, Av. Ej\'ercito 441, Santiago, Chile \label{Arav1}
\and
Millenium Nucleus for Galaxies (MINGAL) \label{Arav2}
\and
IPAC, California Institute of Technology, 1200 E. California Blvd. Pasadena, CA 91125, USA \label{Caltech}
\and
Inter-University Institute for Data Intensive Astronomy (IDIA), Department of Astronomy, University of Cape Town, 7701 Rondebosch, Cape Town, South Africa \label{vac1}
\and
Department of Physics and Astronomy, University of the Western Cape, 7535 Bellville, Cape Town, South Africa \label{vac2}
\and
INAF - Istituto di Radioastronomia, via Gobetti 101, 40129 Bologna, Italy \label{vac3}
\and
Instituto de Alta Investigaci\'on, Universidad de Tarapac\'a, Casilla 7D, Arica, Chile \label{Chile}
\and
School of Physics and Astronomy, Rochester Institute of Technology, 84 Lomb Memorial Drive, Rochester, NY 14623, USA \label{rochester}
\and
Department of Physics, University of California, Santa Barbara, CA 93106, USA \label{SB}
\and
Institut d’Astrophysique Spatiale, Université Paris-Saclay, France \label{spatiale}
\and
European Southern Observatory, Karl-Schwarzschild-Str. 2, 85748 Garching bei München, Germany \label{ESO}
}

\date{Received ; accepted x xx, xxxx}

\abstract{
We present the first deep, targeted LOFAR 144\,MHz source catalogue of the COSMOS field, together with the first 144\,MHz source counts in this field. The data comprise 48\,h of observations. Over the $2.6\,{\rm deg}^{2}$ region adopted for the catalogue analysis, the final Dutch-array image reaches a median rms sensitivity of $\sim100\,\mu$Jy\,beam$^{-1}$ at an angular resolution of $15\arcsec\times6\arcsec$. The resulting catalogue contains 1984 radio sources; a peak S/N $\geq5$ selection is adopted for the resolved/unresolved classification and source-count analysis. We characterised the catalogue using Monte Carlo source-injection and recovery simulations, including dedicated simulations of extended sources, to quantify completeness, flux-density redistribution, and morphology-dependent source-recovery effects. The catalogue reaches 90\% completeness at 1.48\,mJy. The resulting Euclidean-normalised differential source counts are measured down to $\sim$0.4 mJy and show good agreement with LoTSS DR3 144\,MHz source counts and with the VLA-COSMOS 3\,GHz counts scaled to 144\,MHz, with $\sim5\%$ of catalogue sources having multiple 3\,GHz counterparts within a beam-matched search region. Comparison with the VLA-COSMOS 3-GHz catalogue shows good astrometric agreement, with median offsets of $\Delta$RA = $0.35$\arcsec$^{+0.03}_{-0.05}$ and $\Delta$Dec = $-0.83$\arcsec$^{+0.02}_{-0.03}$ (approximately 6\% of the corresponding synthesised-beam dimensions), and no significant radial dependence across the field. The resulting catalogue provides a low-frequency view of COSMOS and a reference dataset for future deeper LOFAR observations and multiwavelength studies of this extensively characterised extragalactic field.
}


\maketitle

\section{Introduction}
\label{sec:intro}

The Cosmic Evolution Survey \citep[COSMOS;][]{scoville2007} is one of the most extensively studied extragalactic fields and was designed to investigate the formation and evolution of galaxies as a function of both cosmic time and large-scale environment. Covering approximately $2\,{\rm deg}^{2}$, the COSMOS field combines deep observations across the electromagnetic spectrum, from X-rays to radio wavelengths, obtained with a wide range of ground- and space-based facilities. This exceptional multiwavelength coverage has enabled detailed studies of galaxy evolution, star formation, active galactic nuclei (AGN), and the role of environment over a broad range of redshifts and physical scales \citep[e.g.][]{scoville2007,Ilbert2009,Laigle2016,Davidzon2017,Darvish2017,Gozaliasl2019,Weaver2022}. The availability of extensive spectroscopy, high-quality photometric redshifts, and measurements of stellar masses and star-formation rates has established COSMOS as a benchmark field for extragalactic studies \citep[e.g.][]{Lilly2007,Laigle2016,Weaver2022,Khostovan2026}. The field is also of continuing importance for major current and next-generation extragalactic surveys, including {\em Euclid} and {\em Nancy Grace Roman} space telescopes.

Deep radio continuum observations provide a complementary view of the galaxy population, tracing both star formation and accretion onto supermassive black holes while being largely unaffected by dust obscuration. COSMOS has been extensively observed at GHz frequencies, most notably through the VLA-COSMOS surveys \citep{Schinnerer2007,Schinnerer2010,Smolcic2017a}, which have provided sensitive, high-resolution radio catalogues and enabled detailed studies of the evolution of radio AGN and star-forming galaxies \citep[e.g.][]{Smolcic2009,Novak2017,Delvecchio2017}. More recently, the MIGHTEE survey with the radio telescope MeerKAT has provided deep continuum and spectral-line observations over the field \citep{Jarvis2016,Heywood2022,Heywood2024,Hale2025}, complemented by multiwavelength counterpart and value-added catalogues \citep{Whittam2024,Hale2026}. The superMIGHTEE project has further extended this radio coverage to lower frequencies, providing deep uGMRT imaging of COSMOS at 650 MHz \citep{Lal2025}. Extending this deep radio coverage to low frequencies is particularly valuable because synchrotron emission at $\sim150$\,MHz probes a complementary part of the radio spectral energy distribution to that sampled by the existing GHz-frequency observations and is particularly sensitive to steep-spectrum and extended emission.

The International LOFAR Telescope \citep[ILT;][]{vanHaarlem2013} has transformed this frequency regime by combining high sensitivity with arcsecond-scale angular resolution. The LOFAR Two-metre Sky Survey \citep[LoTSS;][]{Shimwell2017,Shimwell2019,Shimwell2022,Shimwell2026} has mapped a large fraction of the northern sky at 120--168\,MHz, while the LoTSS Deep Fields have demonstrated the scientific potential of substantially deeper 144\,MHz observations in fields with extensive multiwavelength coverage \citep{Tasse2021,Sabater2021,Kondapally2021,Best2023,Bondi2024,Shimwell2025}. These surveys have provided large samples of radio-selected star-forming galaxies and AGN and enabled studies of their evolution, host-galaxy properties, and the relation between low-frequency radio emission and star formation \citep[e.g.][]{CalistroRivera2017,Smith2021,Bonato2021,Kondapally2022,Best2023}. Star-forming galaxies increasingly dominate the faint 150-MHz population below the mJy level \citep{Best2023}, highlighting the importance of deep low-frequency observations for studies of the evolving galaxy population. 

Despite being one of the most extensively studied extragalactic fields for galaxy evolution and supermassive black hole growth, COSMOS has not, however, benefited from comparably deep LOFAR high band antenna (HBA) coverage. Its low declination, $\delta\simeq+2^\circ$, makes the field challenging to observe from the LOFAR core in the Netherlands. The low elevation limits the duration over which the field can be observed at favourable elevations. This results in a strongly elongated projected array configuration, while projection effects also reduce the effective station collecting area and hence the sensitivity. Low-frequency observations are additionally sensitive to direction-dependent ionospheric and instrumental effects, making accurate calibration and imaging particularly demanding \citep[e.g.][]{vanHaarlem2013,Hale2019,Tasse2021}. The resulting synthesised beam is therefore more elongated than for typical LoTSS observations at higher declinations. However, sensitive 144\,MHz observations of COSMOS can add the low-frequency radio information that is currently missing from its extensive multiwavelength coverage and allow direct comparison with the existing GHz radio surveys.

To address this need, the COSMOS field was observed at 144 MHz with LOFAR for a total of 48\,h. Both Dutch and international LOFAR stations participated in the observations. In this paper, we focus exclusively on the Dutch-array data and present the resulting 144\,MHz image and source catalogue. We describe the calibration and imaging, source extraction using a two-pass {\sc PyBDSF} procedure \citep{Shimwell2025,Shimwell2026}, and characterisation of the resulting catalogue. Its completeness is quantified using source-injection and recovery simulations, while its astrometric performance is assessed through comparison with the VLA-COSMOS 3\,GHz catalogue. We additionally derive the Euclidean-normalised differential source counts and compare them with previous measurements at 144\,MHz and 3\,GHz. The international-baseline data and the corresponding sub-arcsecond imaging will be presented in a separate paper.

This paper is organised as follows. Section~\ref{sec:data} describes the observations, data processing, imaging, and noise properties of the final 144\,MHz image. Section~\ref{sec:pybdsf} describes the source extraction and catalogue construction, with the catalogue columns listed in Appendix~\ref{app:catalogue}. In Section~\ref{sec:quality}, we assess the catalogue completeness and reliability using source-injection simulations and the negative-image test, and examine the astrometric accuracy through comparison with the VLA-COSMOS 3\,GHz catalogue. The 144\,MHz Euclidean-normalised differential source counts and their comparison with previous radio surveys are presented in Section~\ref{sec:source_counts}. Finally, in Section~\ref{sec:conclusion} we summarise our main conclusions and look ahead to deeper observations of COSMOS at low radio frequencies. Additional tests of the catalogue completeness, flux-density scale, astrometry, source blending, and the resolved/unresolved classification are provided in Appendices~\ref{app:completeness_tests} and \ref{app:cat_tests}, while Appendix~\ref{app:catalogue} describes the columns of the released 144\,MHz source catalogue.

Throughout this work, we assume a flat $\Lambda$CDM cosmology with $H_0=70$\,km\,s$^{-1}$\,Mpc$^{-1}$, $\Omega_{\rm m}=0.3$, and $\Omega_{\Lambda}=0.7$. We adopt the radio spectral-index convention $S_{\nu}\propto\nu^{-\alpha}$, where $S_{\nu}$ is the flux density at frequency $\nu$ and $\alpha$ is the spectral index. If not explicitly stated otherwise, $\alpha=0.7$ is assumed.

\section{Data}
\label{sec:data}

\subsection{Observations and observing strategy}

The COSMOS field was observed at 120--168\,MHz with the LOFAR HBA Dutch array (core and remote stations, with baselines up to $\sim120$\,km) and 14 international stations under DDT19\_002 (PI: Vardoulaki). A single pointing was used to cover the COSMOS field, with the catalogue analysis restricted to the central $2.6~{\rm deg}^{2}$ covered by the VLA-COSMOS 3 GHz survey. The observations consisted of twelve 4-hour blocks obtained between 28 January and 3 May 2023. The pointing was centred at (J2000) RA $=150.11916667^\circ$, Dec $=+2.20583333^\circ$. Owing to the low declination of COSMOS, the field reaches a maximum elevation of only $\sim35^\circ$ when observed with LOFAR. We adopted a minimum elevation of $20^\circ$ to avoid the most extreme low-elevation observing conditions, restricting each observing block to approximately 4\,h. In this paper, we present the processing and analysis of the Dutch-array observations using the standard LoTSS processing pipeline described below. The pre-processed data are stored in the LOFAR Long Term Archive (LTA\footnote{\url{https://lta.lofar.eu/}}).

\subsection{Data processing and imaging}
\label{sec:processing}

\begin{figure}
    \centering
    \includegraphics[width=\columnwidth]{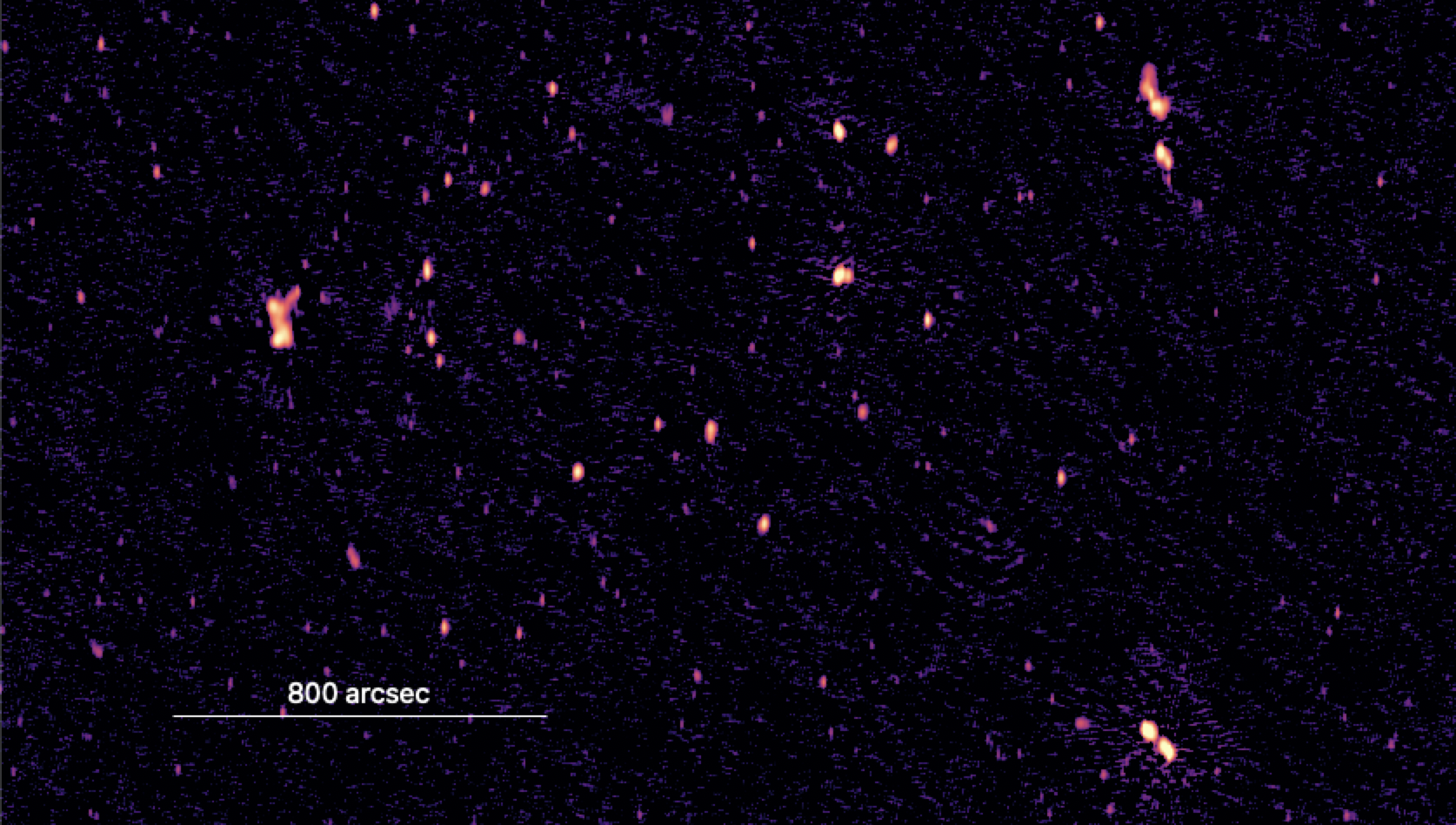}
    \caption{
    Cutout from the 144\,MHz LOFAR image from the COSMOS field, illustrating the typical image quality, noise characteristics, and variety of radio source morphologies present in the survey, including compact sources and resolved radio galaxies with extended emission. The colour scale uses a logarithmic stretch between $0.1$ and $10$\,mJy\,beam$^{-1}$.
    }
    \label{fig:field_cutout}
\end{figure}

The COSMOS observations were processed using the standard LoTSS data reduction pipeline, following the strategy developed for the LoTSS Deep Fields and LoTSS-DR3 \citep{Tasse2021,Shimwell2022,Shimwell2026}. The processing consists of an initial direction-independent calibration using the LOFAR Initial Calibration (LINC) pipeline \citep{vanWeeren2016,Williams2016,deGasperin2019}, followed by direction-dependent calibration and imaging with the DDF-pipeline\footnote{\url{https://github.com/mhardcastle/ddf-pipeline}} \citep{Shimwell2019,Tasse2021}. The LINC pipeline derives the instrumental calibration solutions from dedicated calibrator observations, applies these to the target field, corrects for ionospheric Faraday rotation, flags radio-frequency interference, and removes contamination from bright off-axis radio sources where required. The DDF-pipeline subsequently performs iterative direction-dependent self-calibration to correct for spatially varying ionospheric and primary beam effects, producing high-fidelity wide-field images through facet-based calibration and imaging.

Following the LoTSS processing strategy described by \cite{Shimwell2026}, the twelve 4-hour observations were initially processed independently with LINC and subsequently processed together with the DDF-pipeline. One observation was used to establish the reference calibration and sky model, which was subsequently transferred to the remaining observations via a joint selfcal step, avoiding the need to construct an independent sky model from scratch for each observing block.

The calibrated visibilities from all twelve observing blocks were jointly imaged to produce a single deep continuum image at a central frequency of 144\,MHz. The imaging accounts for the time-, frequency-, and baseline-dependent LOFAR primary beam response, while the final image is corrected for the average primary beam attenuation. Owing to the low declination of the COSMOS field ($\delta \simeq +2^\circ$), the synthesised beam is elongated in the north--south direction compared to typical LoTSS observations, reflecting the reduced projected north--south baseline coverage at transit. Despite the elongated synthesised beam, both compact and extended radio sources are recovered across the field. Figure~\ref{fig:field_cutout} shows a representative cutout from the final 144\,MHz mosaic, illustrating the image quality and the diversity of radio sources, ranging from compact unresolved objects to extended radio galaxies with diffuse emission.

\subsection{Image sensitivity and noise properties}
\label{sec:rms}

\begin{figure}
    \centering
    \includegraphics[width=\columnwidth]{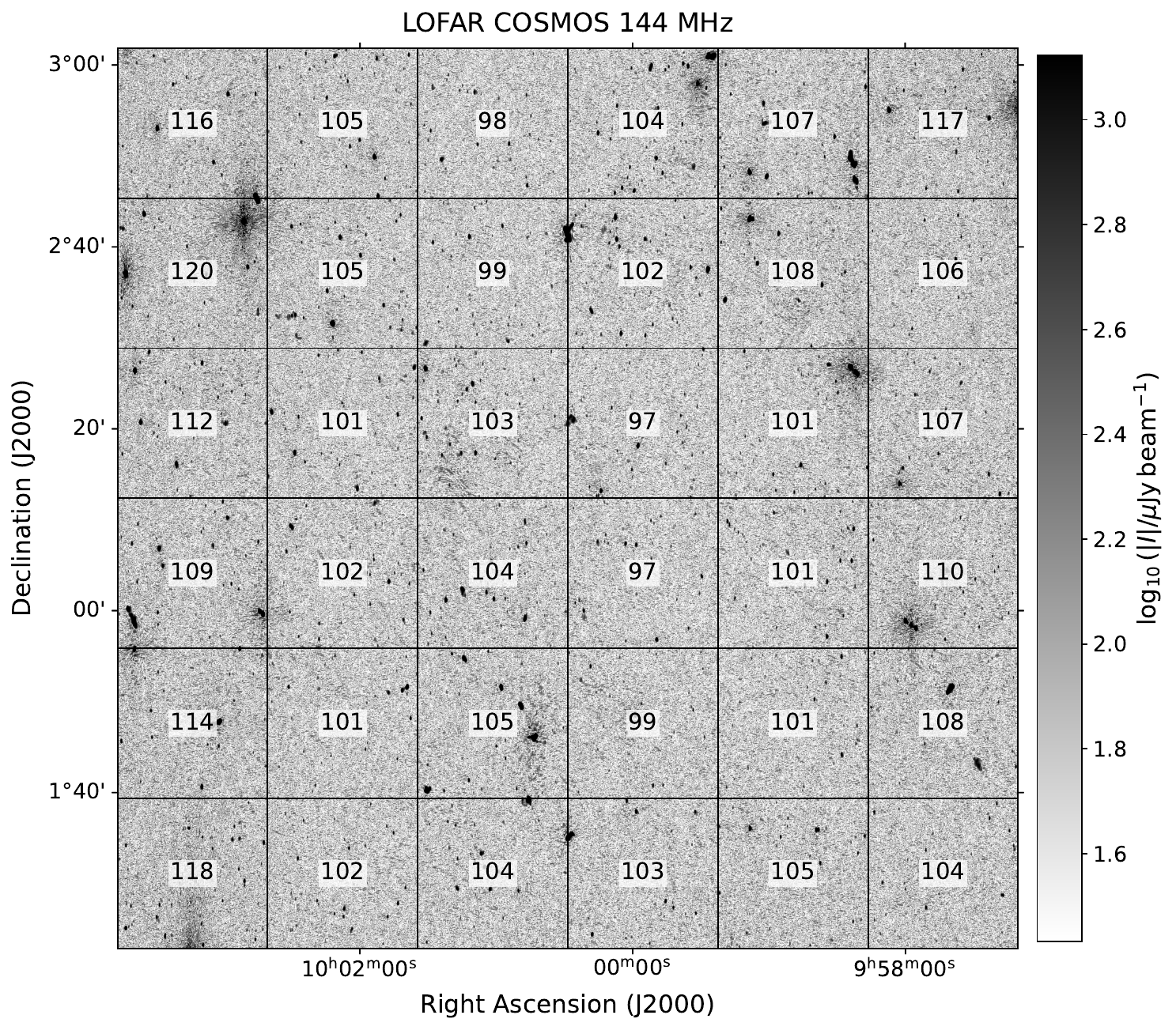}
   \caption{LOFAR COSMOS 144-MHz intensity image over the region covered by the VLA-COSMOS 3\,GHz data, shown in logarithmic greyscale with the colour scale representing $\log_{10}(|I|/\mu{\rm Jy\,beam}^{-1})$. The field is divided into a $6\times6$ grid, and the value shown in each cell is the median local rms derived from the {\sc PyBDSF} rms map, in units of $\mu{\rm Jy\,beam}^{-1}$. This representation is used to quantify the spatial uniformity of the noise across the field, in particular over the central region covered by the VLA-COSMOS 3-GHz data. The rms is lowest across the central part of the field, with typical values of $\sim97$--$105\,\mu{\rm Jy\,beam}^{-1}$, and increases towards parts of the image boundary, reaching $\sim120\,\mu{\rm Jy\,beam}^{-1}$. The median rms within the central $0.5\,{\rm deg}^{2}$ is $99.6\,\mu{\rm Jy\,beam}^{-1}$.}
\label{fig:rms_spatial}
\end{figure}

The final image reaches an angular resolution of 15$\arcsec \times$6$\arcsec$ (PA = 0$^\circ$), estimated from initial PSF fits and subsequently fixed during imaging, as required by the DDF imaging procedure, with a median rms sensitivity of approximately $100\,\mu{\rm Jy\,beam}^{-1}$. To characterise the spatial variation in sensitivity, we use the local rms map produced by {\sc PyBDSF} \citep{pybdsf}. Within the central $0.5\,{\rm deg}^{2}$, defined as a circular region of radius $0.399^\circ$ centred on the pointing centre, the median rms is $99.6\,\mu{\rm Jy\,beam}^{-1}$, with a $16^{\rm th}$--$84^{\rm th}$ percentile range of $95.0$--$106.5\,\mu{\rm Jy\,beam}^{-1}$. Across the $2.6\,{\rm deg}^{2}$ region adopted for the catalogue analysis, the median rms increases to $104.6\,\mu{\rm Jy\,beam}^{-1}$, with a corresponding $16^{\rm th}$--$84^{\rm th}$ percentile range of $98.5$--$114.8\,\mu{\rm Jy\,beam}^{-1}$. As shown in Figure~\ref{fig:rms_spatial}, the cell medians range from approximately $97$ to $120\,\mu{\rm Jy\,beam}^{-1}$, showing that the sensitivity is relatively uniform across most of the 3\,GHz coverage, with a modest increase in the noise towards parts of the image boundary.

\section{Source Extraction and Catalogue}
\label{sec:pybdsf}

Source extraction was performed using a two-pass {\sc PyBDSF} procedure on the final primary-beam-corrected 144\,MHz mosaic in order to optimise the estimation of the local rms noise. We adopted a two-pass PyBDSF strategy based on the refined source-extraction procedure developed for the LoTSS Deep Fields \citep{Shimwell2025,Shimwell2026}, in which an improved rms map is derived after removing faint source emission and subsequently supplied to PyBDSF two-pass for the final source extraction. We describe the full process below. Since the scientific analysis presented in this paper is restricted to the region with extensive ancillary data, source detection was carried out only within the central 2.6\,deg$^{2}$ region, selected to match the 3\,GHz VLA coverage \citep{Smolcic2017a} for follow-up science.

In the first pass, {\sc PyBDSF} was run using the standard LoTSS configuration to generate an initial source catalogue, as well as residual and model images. To construct the rms map used for the second pass, emission associated with detected sources was removed, except for sources with peak flux densities exceeding $150\sigma$, whose residual structure was retained to represent the elevated local noise around bright sources. The resulting rms map was supplied to {\sc PyBDSF} for source extraction from the original image. The $150\sigma$ threshold therefore applies only to the construction of the rms map and not to the final source catalogue. This two-pass procedure increases the number of sources by approximately 30\% compared with a single-pass extraction.

The final source catalogue (see Appendix~\ref{app:catalogue}) was generated using an island detection threshold of \texttt{thresh\_isl}=3 and a peak detection threshold of \texttt{thresh\_pix}=4. The latter is lower than the $4.5\sigma$ threshold adopted in recent LoTSS catalogues \citep{Shimwell2025,Shimwell2026}. The lower threshold retains faint candidates, while ${\rm S/N}\geq5$ is adopted for the main analysis, with reliability assessed in Section~\ref{sec:false_detections}. Additionally we used square rms boxes of 150 pixels on a side, sampled at 15-pixel intervals, with adaptive 60-pixel boxes around bright sources using the same 15-pixel step, zero mean maps, and wavelet decomposition to improve the modelling of extended emission. The remaining {\sc PyBDSF}  parameters follow the standard LoTSS configuration. The final catalogue includes 1984 sources. We retain detections down to the adopted {\sc PyBDSF} threshold in the released catalogue to facilitate counterpart-based studies, while adopting the more conservative peak ${\rm S/N}\geq5$ threshold for the resolved/unresolved classification and source-count analysis presented in Sections~\ref{sec:quality}~\&~\ref{sec:source_counts}.

{\sc PyBDSF} is not optimised for associating multiple radio components with their parent sources. Robust source association generally requires multiwavelength host-galaxy information \citep[e.g.][]{Smolcic2017b,Williams2019,Vardoulaki2019,Vardoulaki2021a,Kondapally2021,Hardcastle2023,Whittam2024} and is beyond the scope of this paper. A value-added COSMOS catalogue incorporating this information is presented in Doukas et al. (submitted).

\subsection{Catalogue properties}
\label{sec:catalogue_properties}

\begin{figure}
    \centering
    \includegraphics[width=\linewidth]{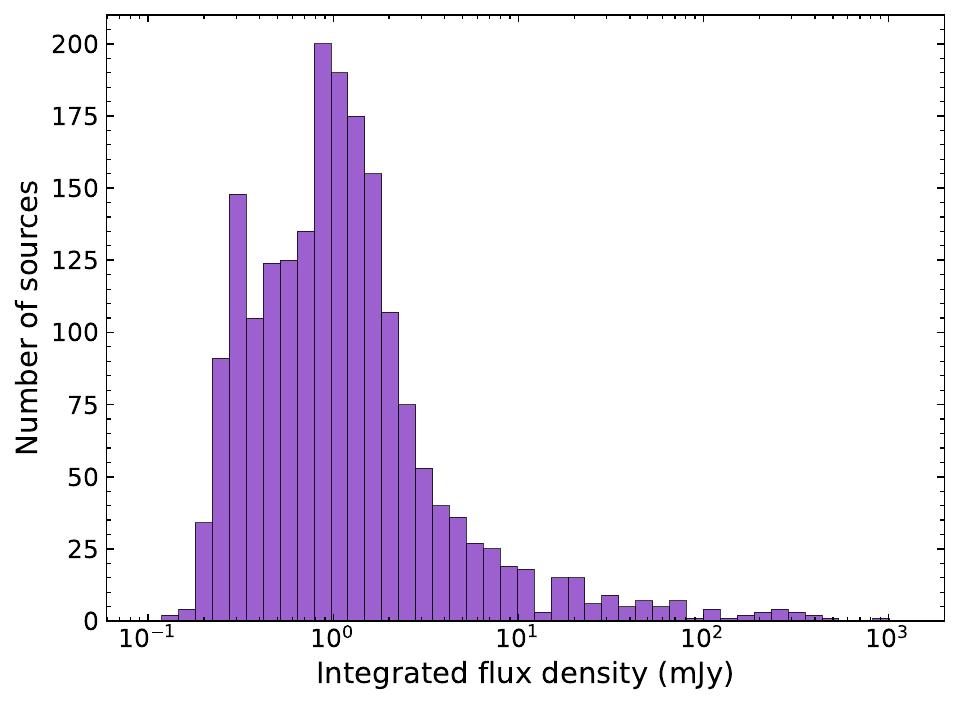}
    \hfill
    \includegraphics[width=\linewidth]{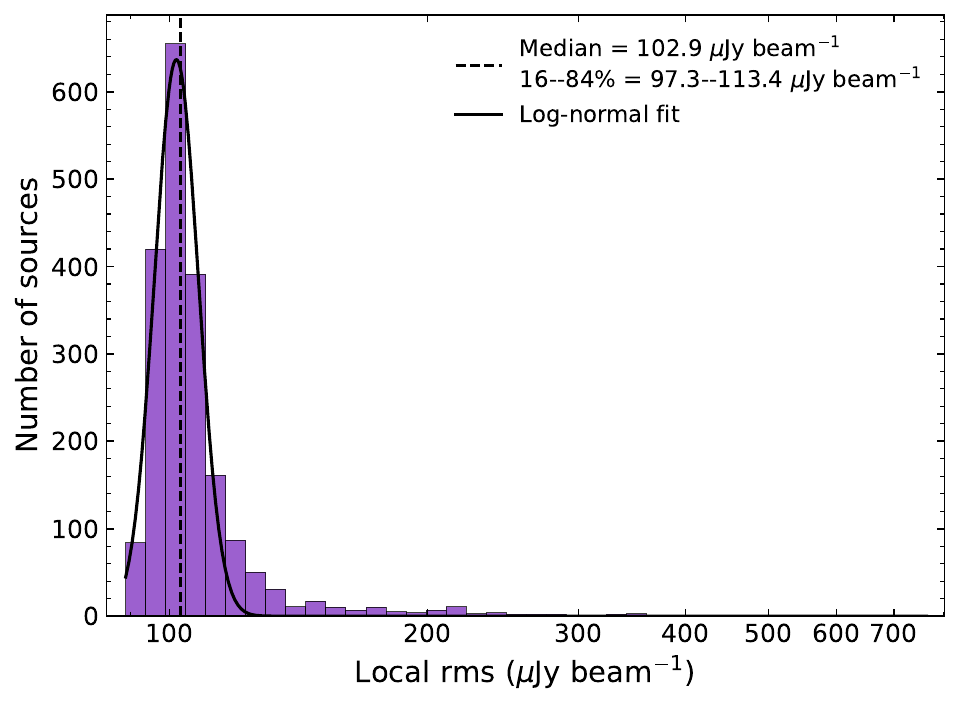}
    \caption{
    Basic properties of the 144\,MHz COSMOS source catalogue for the 2.6~$\rm deg^{2}$ region. Top: distribution of integrated flux densities for all catalogue sources. Bottom: distribution of the local rms noise at the positions of the 1984 catalogued sources measured by {\sc PyBDSF}. The dashed line indicates the median rms of $\sim$100 $\mu$Jy~beam$^{-1}$, with a 16th--84th percentile range of $97.3$--$113.4~\mu$Jy~beam$^{-1}$. The solid black curve shows a log-normal fit to the local-rms distribution.
    }
    \label{fig:catalogue_properties}
\end{figure}

The final {\sc PyBDSF}  catalogue contains 1984 radio sources detected in the central 2.6\,deg$^2$ of the COSMOS field. Figure~\ref{fig:catalogue_properties} summarises two basic properties of the catalogue: the distribution of integrated flux densities and the distribution of the local rms noise measured by {\sc PyBDSF}.

The top panel of Figure~\ref{fig:catalogue_properties} shows the distribution of integrated flux densities. The catalogue has a median integrated flux density of $0.99$\,mJy and spans flux densities from $0.096$\,mJy to $1.25$\,Jy. As expected for a deep radio survey, the number of detected sources increases rapidly towards lower flux densities, before turning over at the faint end where incompleteness becomes important. The small feature at $\sim0.3$\,mJy is dominated by faint sources for which source fitting yields $S_{\rm T}<S_{\rm P}$; in the 0.27--0.34\,mJy bin, 97\% of the sources have $S_{\rm T}/S_{\rm P}<1$. The origin of this feature is not fully understood, but it may be related to source-fitting and selection effects near the detection threshold. The feature disappears when peak rather than integrated flux densities are considered, confirming that it results from the low-S/N flux-fitting rather than from the underlying source distribution.

The bottom panel of Figure~\ref{fig:catalogue_properties} shows the distribution of local rms values at the positions of the sources. The distribution is concentrated around $100\,\mu$Jy\,beam$^{-1}$, with a higher-rms tail associated with noisier regions and artefacts near bright sources.

\subsection{Integrated to peak flux densities}
\label{sec:fluxratios}

\begin{figure}
    \centering
        \resizebox{\hsize}{!}
 {
\includegraphics[width=\linewidth]{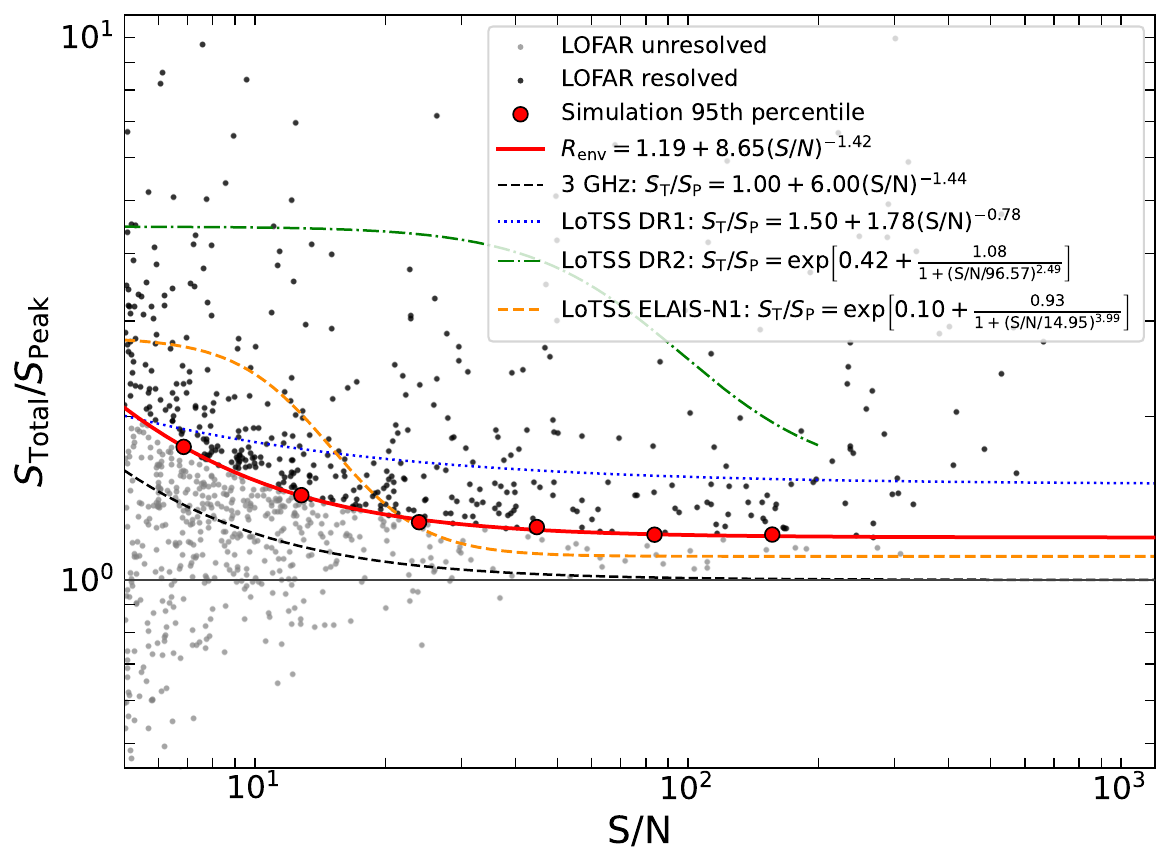}}
 \resizebox{\hsize}{!}
 {
\includegraphics[width=\linewidth]{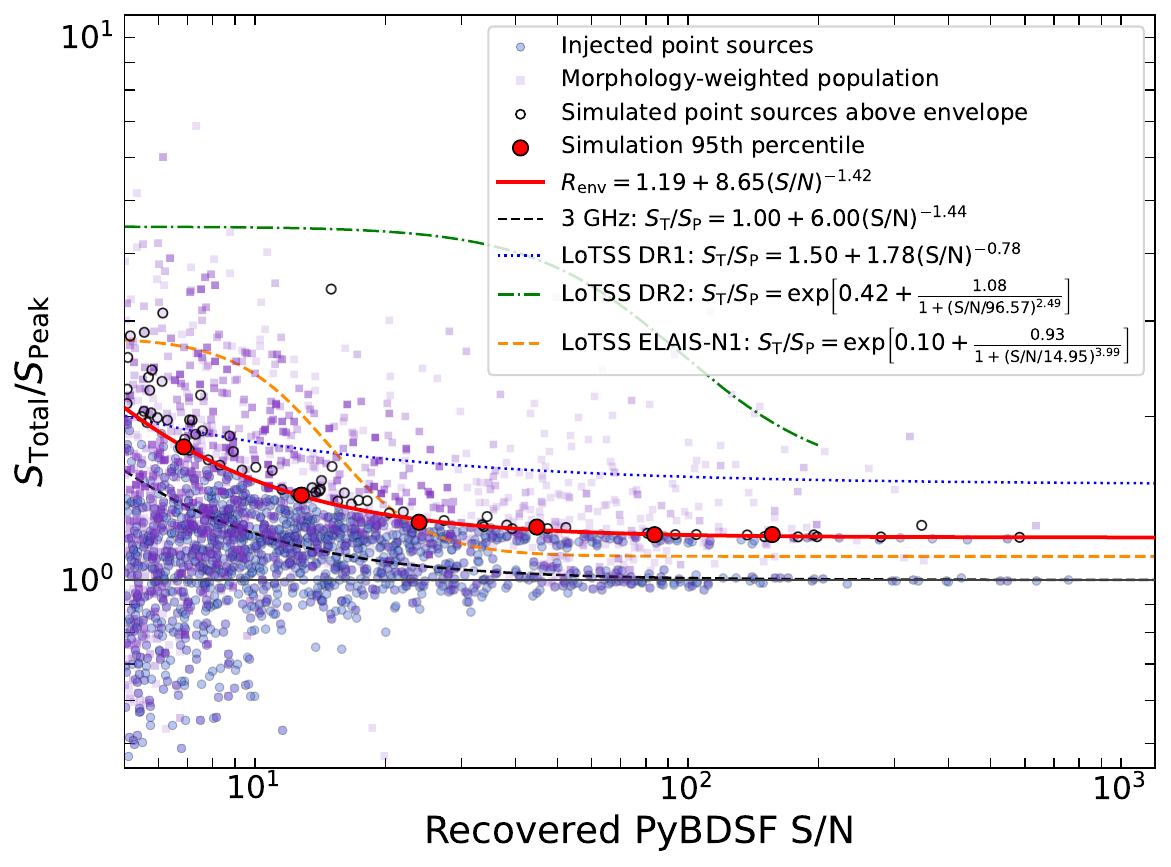}
}  
\caption{Ratio of integrated to peak flux density, $S_{\rm T}/S_{\rm P}$, as a function of peak signal-to-noise ratio. Top: sources in the 144\,MHz COSMOS catalogue, classified as unresolved (grey) or resolved (black) using the simulation-derived envelope. Bottom: recovered injected point sources used to derive the envelope and the morphology-weighted simulated population used for the source-count corrections (Section~\ref{sec:completeness}). Open circles indicate injected point sources recovered above the adopted envelope. In both panels, the red points show the 95th percentiles of the recovered point-source distribution and the red solid line shows the fitted envelope, $R_{\rm env}=1.19+8.65({\rm S/N})^{-1.42}$. The empirical relations adopted for VLA-COSMOS 3\,GHz \citep{Smolcic2017a}, LoTSS DR1 \citep{Shimwell2019}, and LoTSS DR2 \citep{Shimwell2022}, and the LoTSS ELAIS-N1 Deep Field \citep{Shimwell2025} are shown for comparison; the LoTSS DR2 relation is shown only over the S/N range for which it was derived. In both panels, the horizontal dark-grey line marks $S_{\rm T}/S_{\rm P}=1$.
}
    \label{fig:flux_snr}
\end{figure}

In this section we analyse the flux-density measurements of the source catalogue produced using the two-pass {\sc PyBDSF}  procedure described in Section~\ref{sec:pybdsf}. Figure~\ref{fig:flux_snr} shows $S_{\rm T}/S_{\rm P}$ as a function of peak S/N, with the threshold adopted for the simulation-based resolved/unresolved classification and source-count analysis (Section~\ref{sec:source_counts}). For unresolved sources, $S_{\rm T}/S_{\rm P}$ is expected to scatter around unity, with increasing dispersion towards low S/N. This selection does not alter the {\sc PyBDSF}  detection threshold of the catalogue.

To derive an empirical classification boundary appropriate for the COSMOS 144\,MHz data, we use the recovered population of injected point sources (see Section~\ref{sec:completeness}) . Only successfully recovered sources with peak ${\rm S/N}\geq5$ are retained. The recovered point sources are divided into approximately equally populated S/N bins, and the 95th percentile of the $S_{\rm T}/S_{\rm P}$ distribution is measured in each bin. Following the approach of \citet{Smolcic2017a}, these percentiles are fitted with a three-parameter function of the form $R_{\rm env} = \frac{S_{\rm T}}{S_{\rm P}} = C + A({\rm S/N})^{B}$.

These simulations reproduce the effects of the local image noise and the associated uncertainties in the {\sc PyBDSF} source
fitting, but the injected point sources themselves are not subject to residual ionospheric or calibration-related broadening. This is relevant to the real COSMOS sources: the apparently compact and isolated source population has a median $S_{\rm T}/S_{\rm P}=1.41$, indicating additional broadening relative to ideal point sources.

We tested the sensitivity of the resulting classification to the adopted functional form and fitting procedure; these tests and the corresponding alternative envelopes are presented in Appendix~\ref{app:envelope_function}. The adopted fit is $R_{\rm env} = 1.19 + 8.65\,({\rm S/N})^{-1.42}$.

This relation defines the upper envelope of the recovered point-source distribution. For the resolved/unresolved classification, we consider the 906 catalogue sources with ${\rm S/N}\geq5$. Sources lying below the envelope are classified as unresolved, while those lying above it are classified as resolved. For the simulation-based completeness and source-count analyses, we adopt the integrated flux density for resolved sources and the peak flux density for unresolved sources. The simulation-based approach accounts for the S/N-dependent scatter introduced by the noise properties and source-extraction procedure of the COSMOS observations. It does not, however, capture additional broadening of real compact sources arising from residual calibration, deconvolution, or ionospheric effects, since the simulated sources are injected into the residual image. Of these, 428 (47.2\%) are resolved and 478 (52.8\%) unresolved.

The upper panel of Figure~\ref{fig:flux_snr} shows the resulting classification of sources in the COSMOS 144\,MHz catalogue, while the lower panel shows the recovered injected point sources from which the envelope is derived. As expected for a boundary based on the 95th percentile, a small fraction of intrinsically point-like simulated sources scatter above the adopted envelope. The lower panel also shows the morphology-weighted simulated population used to derive the source-count corrections. This population is not used to define the $S_{\rm T}/S_{\rm P}$ envelope and is described in detail in Section~\ref{sec:completeness}.

For comparison, Figure~\ref{fig:flux_snr} also shows the empirical relations adopted for the VLA-COSMOS 3\,GHz survey \citep{Smolcic2017a} and for LoTSS DR1 \citep{Shimwell2019} and DR2 \citep{Shimwell2022}, together with the relation derived for the LoTSS ELAIS-N1 Deep Field \citep{Shimwell2025}. The LoTSS releases similarly use empirical approaches based on the observed behaviour of apparently compact sources to distinguish resolved and unresolved sources. In the more recent LoTSS methodology, the boundary is determined using survival analysis rather than percentile estimates in bins of S/N \citep{Shimwell2025}. The VLA-COSMOS 3\,GHz relation shows a broadly similar S/N dependence to that derived here, but lies somewhat below the COSMOS 144\,MHz envelope. The LoTSS DR1 relation lies above our envelope, while the LoTSS DR2 relation lies below it over most of the S/N range shown. The ELAIS-N1 relation also lies above the COSMOS envelope, particularly at lower S/N, and approaches a nearly constant $S_{\rm T}/S_{\rm P}$ ratio at high S/N. These differences likely reflect the different angular resolutions, sensitivities, noise properties, and source-extraction procedures of the respective surveys.

We also applied the survival-function approach of \citet{Shimwell2025} and examined the radial dependence of the broadening of apparently compact sources. The survival-function diagnostic gives a resolved fraction of 57.4\%, compared with 58.4\% for the adopted envelope, while no significant radial dependence is found. We therefore retain the COSMOS-specific simulation-derived envelope. These tests are described in Appendix~\ref{app:source_broadening}.

\section{Image and catalogue quality}
\label{sec:quality}

\subsection{Spectral indices and flux-density consistency}
\label{sec:spectral_indices}

We cross-matched the LOFAR catalogue with the VLA-COSMOS 3\,GHz catalogue \citep{Smolcic2017a} using a $2$\arcsec\ matching radius, yielding 980 matches. This conservative matching radius was adopted to minimise ambiguous associations and select sources with closely coincident radio positions for the spectral-index analysis. For these sources, we calculated the spectral index between 144\,MHz and 3\,GHz assuming $S_{\nu}\propto\nu^{-\alpha}$. At 144\,MHz, we adopted the peak flux density for unresolved sources and the integrated flux density for resolved sources, following the same prescription used for the source-count analysis. The resulting distribution has a median $\alpha_{144\,\mathrm{MHz}}^{3\,\mathrm{GHz}}=0.67$, a mean of $0.66$, and a standard deviation of $0.27$, with a 16th--84th percentile range of $0.43$--$0.88$. The median value therefore is consistent with the representative value of $\alpha=0.7$ adopted throughout this paper.

\begin{figure*}
    \centering
    \includegraphics[width=\textwidth]{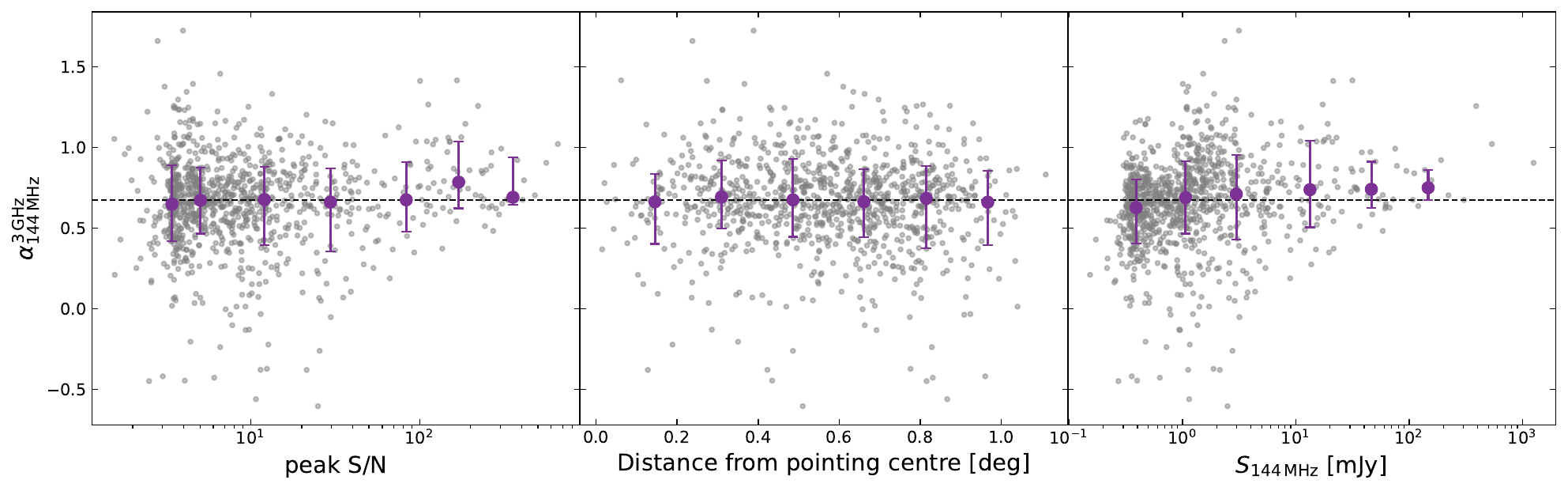}
    
    \caption{
  Spectral indices between the LOFAR 144\,MHz and VLA-COSMOS 3\,GHz flux densities for the 980 cross-matched sources, adopting the convention $S_{\nu}\propto\nu^{-\alpha}$. At 144\,MHz, we adopt the peak flux density for unresolved sources and the integrated flux density for resolved sources. Left: $\alpha_{144\,\mathrm{MHz}}^{3\,\mathrm{GHz}}$ as a function of the LOFAR 144\,MHz peak signal-to-noise ratio. Middle: $\alpha_{144\,\mathrm{MHz}}^{3\,\mathrm{GHz}}$ as a function of distance from the LOFAR pointing centre. Right: $\alpha_{144\,\mathrm{MHz}}^{3\,\mathrm{GHz}}$ as a function of the adopted LOFAR 144\,MHz flux density. Grey circles show individual sources, while the binned purple circles show the median spectral index, with error bars indicating the 16th--84th percentile range. The horizontal dashed line marks the median spectral index of the sample, $\alpha_{144\,\mathrm{MHz}}^{3\,\mathrm{GHz}}=0.67$.
    }
    \label{fig:alpha_plots}
\end{figure*}

Figure~\ref{fig:alpha_plots} shows the spectral indices as a function of LOFAR peak signal-to-noise ratio, distance from the LOFAR pointing centre, and 144\,MHz adopted flux density. The binned values show the median spectral index, with the 16th--84th percentile range illustrating the source-to-source scatter. We find no significant correlation between spectral index and LOFAR signal-to-noise ratio (Spearman $\rho=0.057$, $p=0.06$), or between spectral index and distance from the pointing centre ($\rho=-0.051$, $p=0.10$). In contrast, the spectral index is correlated with 144\,MHz flux density ($\rho=0.239$, $p=3.0\times10^{-14}$), with the binned values showing an apparent flattening towards the faintest flux densities. The interpretation of this trend is limited by the requirement of a 3\,GHz counterpart and by the simple positional matching adopted here, and we therefore do not interpret it as an intrinsic spectral-index--flux-density relation. Increasing the matching radius from $2$\arcsec\ to $3$\arcsec\ increases the number of matches from 980 to 1256, but yields a consistent median spectral index of $\sim 0.68$, with the flux-density correlation remaining significant ($\rho=0.268$, $p=5.0\times10^{-22}$). The derived $\alpha^{3\,\mathrm{GHz}}_{144\,\mathrm{MHz}}$ values are provided in the catalogue presented in Appendix~\ref{app:catalogue}.

\subsection{Catalogue completeness}
\label{sec:completeness}

\begin{figure}
    \centering
    \includegraphics[width=\columnwidth]{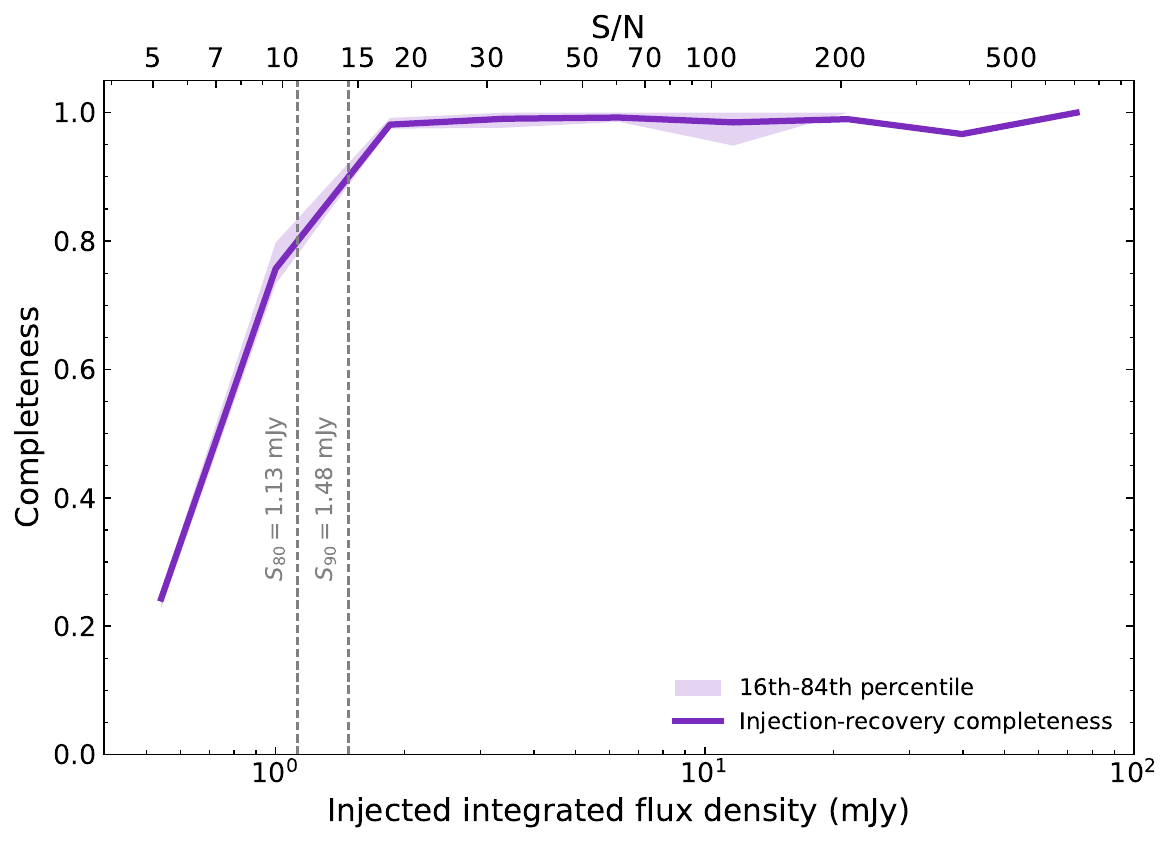}    
        \caption{Completeness as a function of injected flux density, derived from the source-injection and recovery simulations, both intrinsically unresolved and extended sources. Artificial sources were inserted into the residual images, extracted with {\sc PyBDSF} using the same procedure adopted for the COSMOS catalogue, and associated with the input catalogue using a $3^{\prime\prime}$ matching radius. The curve shows the completeness obtained by combining the ten simulation realisations, while the shaded region indicates the $16^{\rm th}$--$84^{\rm th}$ percentile range among the individual realisations. Vertical lines mark the flux densities at which the completeness reaches 80\% and 90\%, $S_{80}=1.13$\,mJy and $S_{90}=1.48$\,mJy, respectively. The completeness increases rapidly around 1\,mJy and approaches unity above approximately 2--3\,mJy.
       }
    \label{fig:completeness}
\end{figure}

To quantify the completeness of the source catalogue, we performed Monte Carlo source-injection and recovery simulations. The simulated sources were injected into the residual images and the resulting images were processed with the same source-extraction procedure adopted for the COSMOS catalogue (Section~\ref{sec:pybdsf}). Ten independent realisations were generated, with 3000 injected sources per realisation, giving a total of 30,000 injected sources. The injected integrated flux densities follow a broken power-law distribution over 0.05--100\,mJy, with $dN/dS\propto S^{-\gamma}$ and $\gamma=1.1$--$2.3$, and include both compact and extended sources to characterise source recovery and the redistribution of measured flux densities introduced by the imaging and source-extraction procedure. We emphasise that these baseline simulations already contain both intrinsically unresolved and extended sources; the additional extended-source simulations described below are used to improve the sampling of the size-dependent recovery response, rather than to introduce an extended-source population absent from the baseline simulations.

The injected catalogues were cross-matched to the recovered {\sc PyBDSF} catalogues using a nearest-neighbour matching radius of $3^{\prime\prime}$. A source was considered successfully recovered when a counterpart was found within this radius and its recovered peak  satisfied ${\rm S/N}\geq5$. We repeated the matching using radii of $2^{\prime\prime}$ and $4^{\prime\prime}$ to test the sensitivity of the results to the adopted association criterion. The results are presented in Appendix~\ref{app:completeness_tests}. The completeness obtained with $3^{\prime\prime}$ and $4^{\prime\prime}$ are effectively identical, and we therefore adopt $3^{\prime\prime}$ throughout the analysis.

We first quantify the \emph{completeness} $C_{\rm rec}=N_{\rm rec}/N_{\rm in}$, where $N_{\rm in}$ is the number of injected sources in a given injected-flux-density bin and $N_{\rm rec}$ is the number satisfying the recovery criteria above. Figure~\ref{fig:completeness} shows the resulting completeness. The central curve is calculated from the total numbers of injected and recovered sources across the ten realisations, while the shaded region represents the $16^{\rm th}$--$84^{\rm th}$ percentile range of the completeness measured in the individual realisations. Thus, the shaded range reflects Monte Carlo variability, not the full systematic uncertainty in completeness. The completeness rises rapidly around 1\,mJy, reaching 80\% at $S_{80}=1.13$\,mJy and 90\% at $S_{90}=1.48$\,mJy, and approaches unity above approximately 2--3\,mJy. Since the simulations use residual images, the rms maps re-derived after source injection may differ from that of the real catalogue, particularly near bright sources. We tested this by repeating one realisation using the rms and mean maps derived from the original science image. The recovery was effectively unchanged (543 versus 542 sources). We also performed a leave-one-out test across the ten realisations. The resulting 80\% and 90\% completeness limits varied by only 1.117--1.137 and 1.473--1.487\,mJy, respectively. We additionally tested the effect of the residual broadening identified in Sect.~\ref{sec:fluxratios} by applying the empirically derived broadening distribution to the point-source component of the simulations. The resulting completeness is similar to that obtained from the original baseline simulations, indicating that residual broadening of the point-source component does not significantly affect the quoted completeness levels.

The correction applied to the differential source counts accounts for measurement uncertainties that redistribute sources between flux-density bins and for the dependence of source detectability and recovered flux density on angular extent. Both effects are quantified by comparing the injected and recovered source populations in the simulations. We therefore follow the input--output approach adopted for the VLA-COSMOS 3\,GHz source counts by \citet{Smolcic2017a}. The input distribution is binned according to the injected integrated flux density, whereas the recovered distribution is constructed using the flux-density estimator adopted for the observed catalogue: integrated flux density for sources classified as resolved and peak flux density for sources classified as unresolved according to the $S_{\rm T}/S_{\rm P}$ envelope derived in Section~\ref{sec:fluxratios}. This procedure accounts simultaneously for source recovery and for the redistribution of sources between flux-density bins.

\begin{table}
\centering
\caption{Catalogue completeness and effective source-count corrections derived from the source-injection simulations. Uncertainties on $f_{\rm corr}$ include the simulation-realisation and morphology-model uncertainties.}
\label{tab:completeness_correction}
\setlength{\tabcolsep}{4pt}
\begin{tabular}{cccccc}
\hline\hline
 \multicolumn{1}{c}{$S_{\rm min}$--$S_{\rm max}$} & $N_{\rm in}$ & $N_{\rm rec}$ & $C_{\rm rec}$ & $C_{\rm eff}$ & $f_{\rm corr}$ \\
 \multicolumn{1}{c}{(mJy)} & & & & & \\
\hline
0.063--0.117 & 7117 & 2 & 0.0003 & \ldots & \ldots \\
0.117--0.215 & 6214 & 7 & 0.001 & \ldots & \ldots \\
0.215--0.398 & 4949 & 105 & 0.021 & \ldots & \ldots \\
0.398--0.736 & 3507 & 853 & 0.243 & 0.359 & $2.786_{-0.120}^{+0.664}$ \\
0.736--1.359 & 2456 & 1859 & 0.757 & 0.525 & $1.904_{-0.058}^{+0.085}$ \\
1.359--2.512 & 1396 & 1370 & 0.981 & 0.931 & $1.074_{-0.060}^{+0.060}$ \\
2.512--4.642 & 740 & 733 & 0.991 & 1.088 & $0.919_{-0.036}^{+0.044}$ \\
4.642--8.577 & 395 & 392 & 0.992 & 1.025 & $0.976_{-0.039}^{+0.043}$ \\
8.577--15.849 & 198 & 195 & 0.985 & 0.904 & $1.106_{-0.035}^{+0.037}$ \\
15.849--29.286 & 99 & 98 & 0.990 & 0.960 & $1.042_{-0.023}^{+0.022}$ \\
29.286--54.117 & 30 & 29 & 0.967 & 0.967 & $1.034_{-0.072}^{+0.080}$ \\
54.117--100.000 & 17 & 17 & 1.000 & 0.941 & $1.062_{-0.062}^{+0.062}$ \\
\hline
\end{tabular}
\end{table}

Because extended sources have lower peak surface brightness than compact sources of the same integrated flux density, their recovery depends on source extent. We therefore used dedicated extended-source injection simulations to characterise the dependence of source recovery and measured flux density on intrinsic source size. Here, morphology refers specifically to the angular size and shape of the injected Gaussian sources, rather than to their physical radio morphology (e.g. the presence of jets or lobes). The parent population was drawn from the intrinsically extended sources in the main simulations, restricted to integrated flux densities of $0.5\leq S_{\rm T}<5$\,mJy. This range was chosen to concentrate the additional simulations towards the faint regime, where source extent has the greatest effect on recovery, rather than to define an upper flux-density limit for the source-count analysis. To improve the sampling of the recovery behaviour as a function of source extent, the selected sources were deliberately broadened: their intrinsic major and minor axes were increased by a factor of 2.5, preserving their axial ratios, and the resulting major-axis sizes were restricted to $2$--$15\arcsec$. The purpose of this scaling was to populate a wider range of angular sizes for measuring the size-dependent recovery response; it does not represent an assumed intrinsic size distribution. Source position angles were drawn uniformly between $0^\circ$ and $180^\circ$. A total of 1500 extended sources were injected over three independent realisations of 500 sources each. The sources were modelled as elliptical Gaussians, convolved with the $15\arcsec\times6\arcsec$ restoring beam, and injected into the same residual images and over the same spatial footprint as the main simulations.

These simulations deliberately oversample sources with larger angular sizes and are therefore not intended to represent the intrinsic angular-size distribution of the COSMOS population. Instead, they provide a response sample with which to characterise the dependence of source recovery and measured flux density on intrinsic source extent. We subsequently weight the extended-source response sample using the observed $S_{\rm T}/S_{\rm P}$ distribution and resolved-source fraction in each flux-density bin, so that the weighted simulations approximately reproduce the observed morphology of the COSMOS source population. Specifically, in four intrinsic flux-density bins between 0.5 and 4.64\,mJy, we model the intrinsic major-axis distribution using a log-normal weighting function of the form $w(\theta)\propto\exp\{-\frac{1}{2}[(\ln\theta-\ln\theta_0)/\sigma_{\ln\theta}]^2\}$, where $\theta$ is the intrinsic major axis and $\theta_0$ is the characteristic angular scale fitted independently in each flux-density bin. These bins follow the logarithmic flux-density binning adopted for the completeness and source-count analysis, restricted to the faint regime targeted by the dedicated extended-source simulations. The log-normal form is used only to weight the simulated source sizes and is not physically motivated. The characteristic size $\theta_0$ is determined independently in each flux-density bin by jointly matching the recovered simulated $S_{\rm T}/S_{\rm P}$ distribution and resolved-source fraction to those measured in the COSMOS catalogue, using the extended-source simulations to model the {\sc PyBDSF} response as a function of intrinsic angular size; the observed source-count amplitudes are not used in this optimisation. The weighted extended-source population replaces the corresponding extended component of the baseline simulations while preserving the expected number of injected sources in each flux-density bin. We adopt $\sigma_{\ln\theta}=0.35$ for the fiducial model and repeat the procedure for $\sigma_{\ln\theta}=0.25$, $0.45$, and $0.55$ to quantify the sensitivity of the source-count corrections to the assumed width of the intrinsic size distribution. The resulting range in correction factors is included as the morphology-model uncertainty. The fiducial morphology-weighted population is then used to derive the final input--output source-count correction.

Although we find no significant variation in $S_{\rm T}/S_{\rm P}$ with distance from the pointing centre, residual source broadening could still affect the source counts by reducing the peak surface brightness of compact sources and hence altering their recovery and flux-density redistribution. We therefore tested the sensitivity of the source-count corrections to this effect (see Appendix~\ref{app:source_broadening}). The intrinsically unresolved component of the baseline simulations does not include residual broadening. We constructed an empirical broadening distribution from the apparently compact and isolated COSMOS sources and applied it to the simulated compact-source population. Since the fraction of intrinsically compact sources subject to residual broadening is not independently constrained, we considered a range $f_{\rm smear}=0$--1, where $f_{\rm smear}=0$ corresponds to the fiducial morphology-informed correction and $f_{\rm smear}=1$ to the limiting case in which the entire compact-source component is subject to the empirical broadening distribution. The intrinsically extended component and the adopted $S_{\rm T}/S_{\rm P}$ resolved/unresolved classification were kept unchanged.

The resulting corrected source counts differ from the fiducial values by less than 5\% in most flux-density bins (Figure~\ref{fig:smearing_counts}). The largest difference occurs in the $0.736$--$1.359$\,mJy bin, where the corrected count increases by approximately 18\% for the limiting $f_{\rm smear}=1$ case. This is driven primarily by the reduction in peak surface brightness of broadened compact sources, which reduces their recovery fraction and alters flux-density redistribution close to the detection threshold. We therefore retain the fiducial morphology-informed corrections for the source counts and treat the range obtained from this test as a separate systematic sensitivity to residual source broadening.

We emphasise that the extended-source simulations are used only to characterise the morphology-dependent response required for the source-count corrections and do not replace the baseline simulations used to estimate the catalogue completeness. The completeness presented in Figure~\ref{fig:completeness} is therefore derived from the main injection--recovery simulations, while the morphology-weighted population is introduced only when deriving the input--output correction applied to the source counts.

We define the corresponding effective completeness as $C_{\rm eff}(S)=N_{\rm out}(S)/N_{\rm in}(S)$ where $N_{\rm in}$ denotes the input source distribution and $N_{\rm out}$ the morphology-weighted recovered distribution in the corresponding measured-flux-density bin. Unlike $C_{\rm rec}$, $C_{\rm eff}$ is not a pure detection probability: because sources can migrate between flux-density bins, it can exceed unity. The correction applied to the observed differential source counts is therefore $f_{\rm corr}(S)=C_{\rm eff}^{-1}(S)=N_{\rm in}(S)/N_{\rm out}(S)$. Consequently, $f_{\rm corr}$ can also be smaller than unity when flux-density redistribution produces a net excess of recovered sources in a given measured-flux-density bin.

Table~\ref{tab:completeness_correction} summarises the completeness and effective source-count corrections. Below 0.398\,mJy the recovery fraction is too small to derive a robust finite correction. The largest correction, $f_{\rm corr}=2.786^{+0.664}_{-0.120}$, occurs in the 0.398--0.736\,mJy bin, which lies close to the detection limit and should therefore be interpreted with caution. The corrections approach unity above approximately 1.4\,mJy.

The uncertainties on $f_{\rm corr}$ include both the variation among the independent simulation realisations and the uncertainty associated with the adopted morphology model. These final correction factors, rather than $1/C_{\rm rec}$, are applied to the differential source counts presented in Section~\ref{sec:source_counts}.

\subsection{False-detection rate}
\label{sec:false_detections}

To assess the reliability of the source catalogue, we estimated the false-detection rate using the negative-image method. The final 144\,MHz image was multiplied by $-1$, and source extraction was repeated using the same detection parameters and $2.6\,{\rm deg}^{2}$ analysis region adopted for the positive image. The rms maps derived from the original positive science image were supplied to the negative-image extraction, ensuring that the positive and negative catalogues were extracted using the same local noise model. Since genuine astronomical sources appear as negative features in the inverted image, positive detections in the negative image provide an empirical estimate of the contamination arising from noise fluctuations and residual imaging artefacts.

\begin{figure}
    \centering
    \includegraphics[width=\columnwidth]{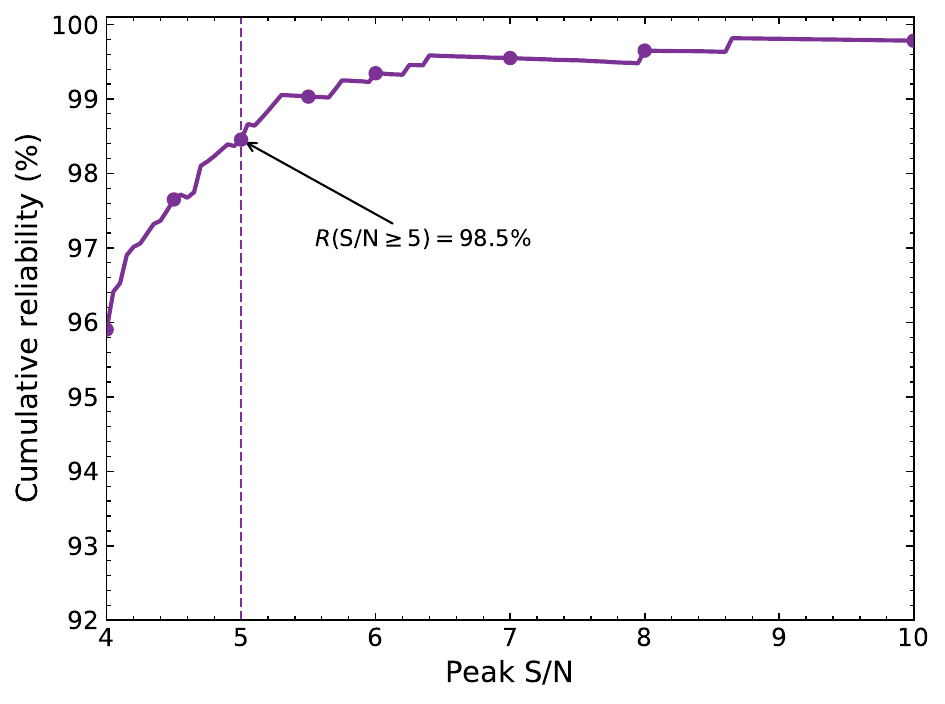}
    \caption{Cumulative catalogue reliability as a function of peak signal-to-noise ratio, derived from source extraction on the inverted 144\,MHz image using the same rms maps as for the positive-image catalogue extraction. The reliability is calculated as $R=1-N_{\rm neg}/N_{\rm pos}$, where $N_{\rm neg}$ and $N_{\rm pos}$ are the cumulative numbers of negative- and positive-image detections above each S/N threshold. The vertical dashed line marks the peak ${\rm S/N}=5$ threshold adopted for the simulation-based resolved/unresolved classification and source-count analysis, at which the inferred cumulative reliability is $98.5\%$. Negative-image detections include both noise fluctuations and residual imaging artefacts.}
    \label{fig:false_detection}
\end{figure}

We compare the cumulative numbers of positive and negative detections above a given peak signal-to-noise ratio. The false-detection fraction is defined as $f_{\rm false}=N_{\rm neg}/N_{\rm pos}$, where $N_{\rm neg}$ and $N_{\rm pos}$ are the numbers of sources detected in the negative and positive images, respectively, above the corresponding S/N threshold. The cumulative catalogue reliability is then $R=1-f_{\rm false}$. The resulting values are given in Table~\ref{tab:false_detection} in Appendix~\ref{app:completeness_tests}, and the cumulative reliability is shown in Figure~\ref{fig:false_detection}. The S/N thresholds considered here refer to the post-extraction ratio of fitted peak flux density to the corresponding island rms and are distinct from the \texttt{thresh\_pix}=4 criterion used by {\sc PyBDSF} during source extraction (Section~\ref{sec:pybdsf}).

At the ${\rm S/N}\geq5$ threshold adopted for the simulation-based resolved/unresolved classification and source-count analysis, 14 negative detections are found compared with 906 positive sources, corresponding to a false-detection fraction of 1.5\% and a catalogue reliability of 98.5\%. The reliability exceeds 99\% at ${\rm S/N}\geq5.5$ and reaches 99.6\% at ${\rm S/N}\geq8$. As a sensitivity test, we also re-derived the rms map from the inverted image; this yields 30 negative detections at ${\rm S/N}\geq5$, corresponding to a reliability of 96.7\%. We do not adopt this estimate because re-deriving the rms after inversion changes the local noise model relative to that used for the positive catalogue, particularly around bright-source residuals.

\subsection{Astrometric accuracy}
\label{sec:astrometry}

We assessed the astrometric accuracy of the LOFAR catalogue by comparing the positions of matched sources with those in the VLA-COSMOS 3\,GHz catalogue, which provides the astrometric reference for the field. The positional offsets were calculated as $\Delta{\rm RA}=({\rm RA}_{144\,{\rm MHz}}-{\rm RA}_{3\,{\rm GHz}})\cos({\rm Dec})$ and $\Delta{\rm Dec}={\rm Dec}_{144\,{\rm MHz}}-{\rm Dec}_{3\,{\rm GHz}}$.

\begin{figure}
    \centering
    \includegraphics[width=\columnwidth]{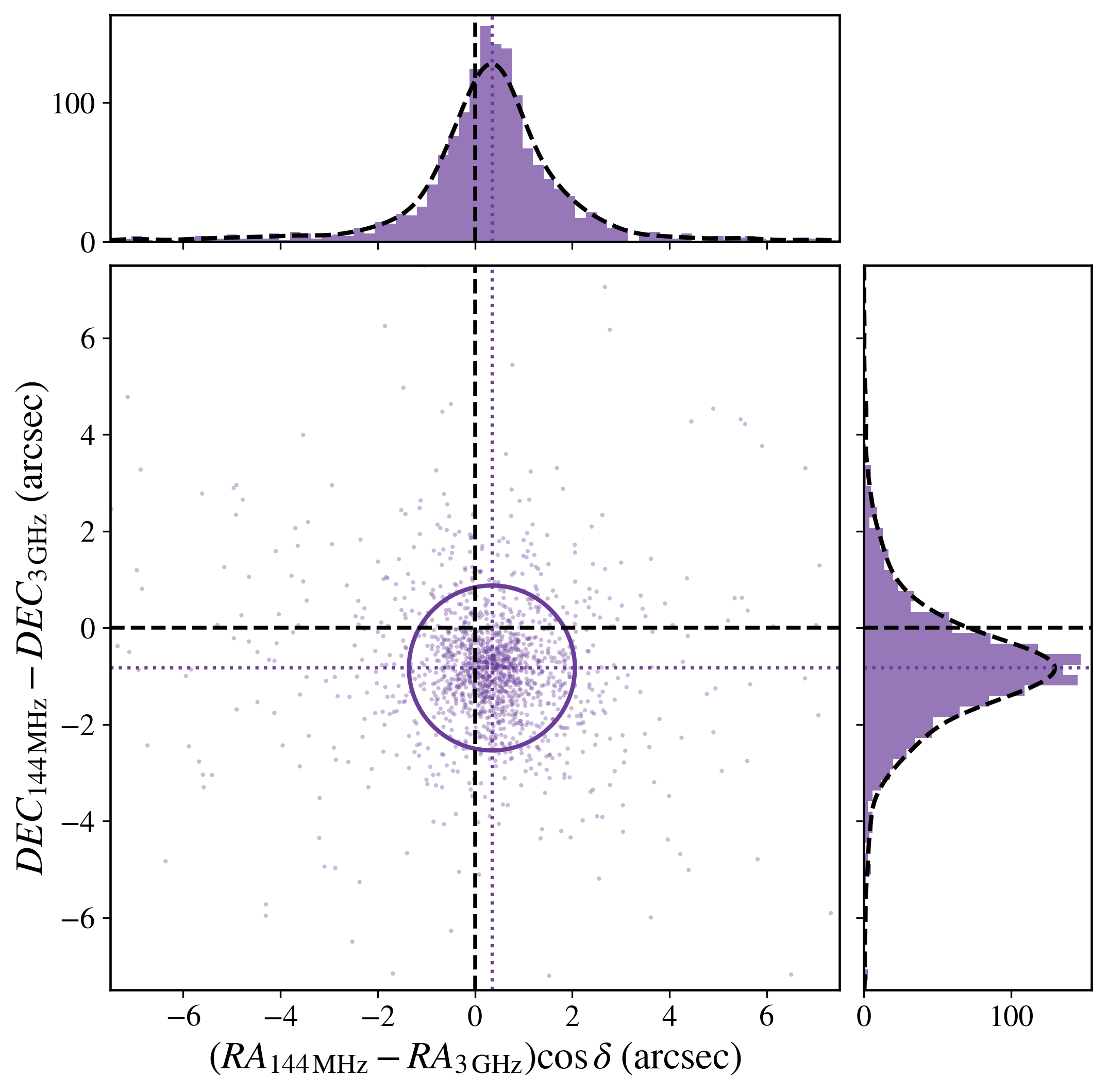}
    \caption{Astrometric offsets between the LOFAR 144\,MHz source positions and their matched VLA-COSMOS 3\,GHz counterparts. The central panel shows the two-dimensional distribution of $\Delta{\rm RA}\cos\delta$ and $\Delta{\rm Dec}$ for sources with $|\Delta{\rm RA}\cos\delta|,|\Delta{\rm Dec}|\leq7.5$\arcsec, corresponding to half the $15$\arcsec\ major-axis FWHM of the LOFAR restoring beam, while the upper and right panels show the one-dimensional distributions. Dashed lines mark zero offset and the dashed curves show kernel-density estimates of the distributions. The circle is centred on the median offset and encloses 68\% of the sources satisfying this positional-offset selection. The median offsets are $\Delta{\rm RA}=0.35$\arcsec$^{+0.03}_{-0.05}$ and $\Delta{\rm Dec}=-0.83$\arcsec$^{+0.02}_{-0.03}$, with uncertainties estimated by bootstrap resampling. The corresponding scatters, estimated using the scaled median absolute deviation, are $0.92$\arcsec\ and $0.93$\arcsec, respectively.}
    \label{fig:astrometry}
\end{figure}

Figure~\ref{fig:astrometry} shows the two-dimensional distribution of the positional offsets and the corresponding one-dimensional distributions. We characterise the astrometric offsets for sources with $|\Delta{\rm RA}|,|\Delta{\rm Dec}|\leq7.5$\arcsec, corresponding to half the $15$\arcsec\ major-axis FWHM of the LOFAR restoring beam. This provides a resolution-motivated window within which to quantify the central distribution of positional offsets. Within this region, we measure median offsets of $\Delta{\rm RA}=0.35$\arcsec$^{+0.03}_{-0.05}$ and $\Delta{\rm Dec}=-0.83$\arcsec$^{+0.02}_{-0.03}$, where the uncertainties on the medians were estimated by bootstrap resampling. These offsets correspond to approximately 6\% of the respective synthesised-beam dimensions. The corresponding scatters, estimated using the scaled median absolute deviation, are 0.92\arcsec in RA and 0.93\arcsec in Dec. The radius enclosing 68\% of the matched sources is 1.71\arcsec. These values reflect the combined positional uncertainties of both catalogues and, for resolved sources, intrinsic differences in the radio emission traced at 144\,MHz and 3\,GHz. Given the low elevation of the COSMOS observations and the correspondingly elongated N–S synthesised beam, a residual calibration or ionospheric contribution to the declination offset cannot be excluded. However, since the astrometric offsets show no significant spatial dependence, we cannot distinguish such a contribution from a small systematic offset between the LOFAR and VLA-COSMOS astrometric reference frames.

\begin{figure}
    \centering
    \includegraphics[width=\columnwidth]{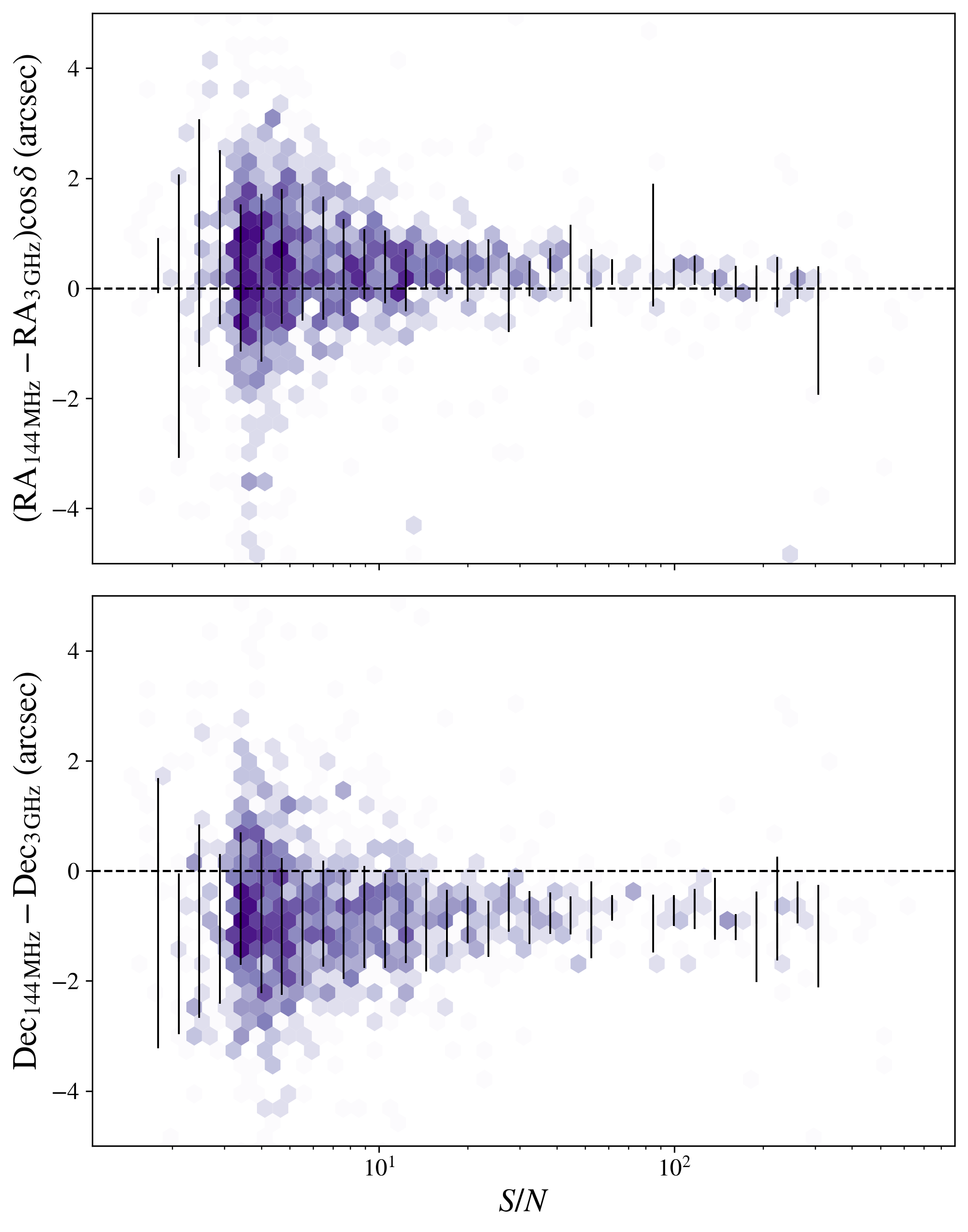}
    \caption{Astrometric offsets as a function of peak signal-to-noise ratio for the 1470 LOFAR sources matched to VLA-COSMOS 3\,GHz counterparts and satisfying $|\Delta{\rm RA}\cos\delta|,|\Delta{\rm Dec}|\leq7.5$\arcsec. The upper panel shows
$(\mathrm{RA}_{144\,\mathrm{MHz}}-\mathrm{RA}_{3\,\mathrm{GHz}})\cos\delta$ and the lower panel shows $\mathrm{Dec}_{144\,\mathrm{MHz}}-\mathrm{Dec}_{3\,\mathrm{GHz}}$. The purple hexagonal bins show the source density in the offset--S/N plane. Black markers show the median offset in bins of S/N, and vertical bars indicate the corresponding $16^{\rm th}$--$84^{\rm th}$ percentile ranges. Horizontal dashed lines mark zero offset.     } 
    \label{fig:astrometry_snr}
\end{figure}

We further examined whether the positional offsets depend on source signal-to-noise ratio. Figure~\ref{fig:astrometry_snr} shows $\Delta{\rm RA}$ and $\Delta{\rm Dec}$ as a function of peak S/N. The scatter is largest close to the catalogue detection threshold and decreases steadily towards higher S/N, as expected from the improved accuracy of brighter sources. The median right ascension offset approaches zero at moderate and high S/N, while the declination offset remains systematically negative, at approximately $-1\arcsec$, over most of the sampled S/N range. The decreasing scatter in the positional offsets with increasing S/N is expected, since the uncertainty in the measured position of a source scales approximately inversely with its signal-to-noise ratio. The persistence of the declination offset at high S/N demonstrates that it is not dominated by measurement noise, but instead reflects a small systematic offset.

To assess the effect of source morphology on the measured positional offsets, we repeated the analysis using only the 425 sources classified as unresolved according to the $S_{\rm T}/S_{\rm P}$ criterion of Section~\ref{sec:fluxratios}. The median offsets remain similar, 
$\Delta{\rm RA}=+0.39\arcsec$ and $\Delta{\rm Dec}=-0.78\arcsec$, while the scatter decreases to $0.59\arcsec$ and $0.73\arcsec$, respectively. This indicates that extended-source morphology contributes to the positional scatter but does not drive the systematic astrometric offset (Appendix~\ref{app:astrometry_unresolved}).

We additionally find no significant dependence of the astrometric offsets on distance from the pointing centre, either for the full matched sample or for sources with ${\rm S/N}\geq20$ (Appendix~\ref{app:astrometry_unresolved}), indicating no evidence for a field-dependent astrometric distortion. We provide astrometrically corrected RA and Dec coordinates in the released catalogue by applying the measured median offsets relative to the VLA-COSMOS 3\,GHz reference frame. The original LOFAR positions are retained in separate columns for reproducibility.

\section{Euclidean-normalised source counts}
\label{sec:source_counts}

\begin{figure}
    \centering
    \includegraphics[width=\columnwidth]{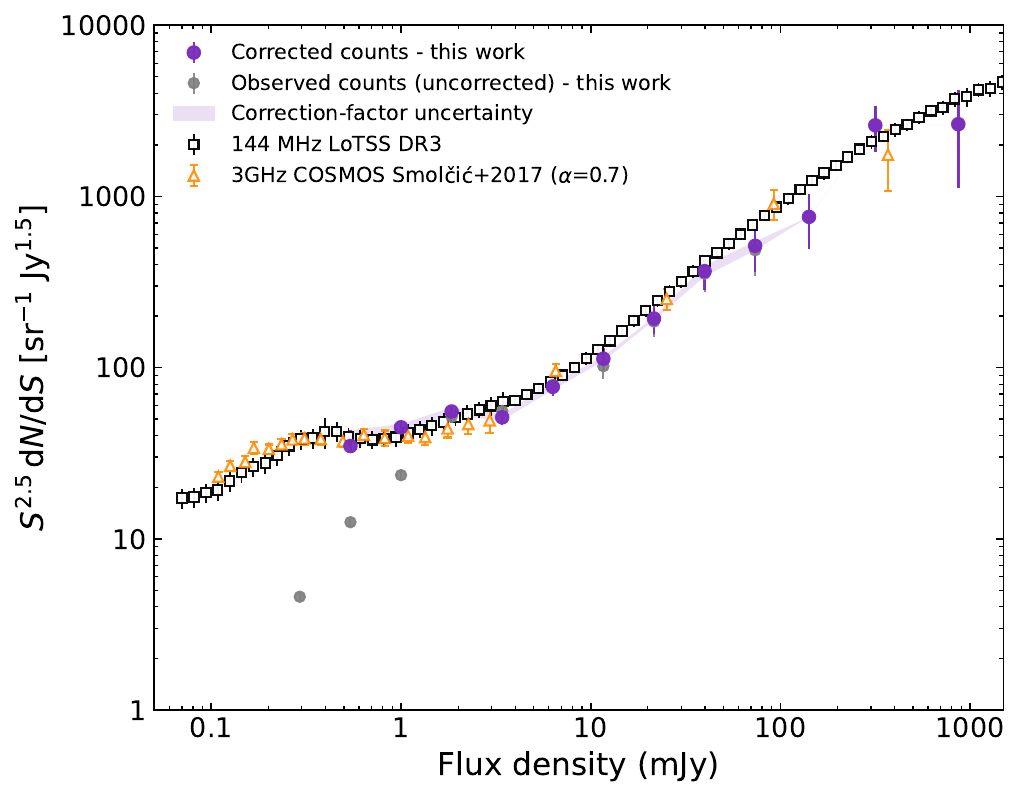}
   \caption{Euclidean-normalised differential source counts at 144 MHz. Grey points show the raw COSMOS counts and purple points the counts after applying the flux-density-dependent source-count corrections derived in Section~\ref{sec:completeness}. Error bars show Poisson uncertainties propagated through the source-count correction, while the purple shaded region shows the additional 16th--84th percentile correction-factor uncertainty. Black open squares show the LoTSS DR3 source counts from \citet{Shimwell2026}, and orange open triangles the VLA-COSMOS 3 GHz counts from  \citet{Smolcic2017a}, scaled to 144 MHz assuming $S_{\nu}\propto\nu^{-0.7}$.
}
    \label{fig:sourcecounts}
\end{figure}

We derived the differential source counts at 144 MHz for the 2.6\,deg$^{2}$ COSMOS region, using the flux-density prescription defined in Section~\ref{sec:completeness} for both the observed and simulated populations. We use the simulation bins at the faint end and broader bins at higher flux densities to reduce small-number statistics.

For each flux-density bin, the differential source counts were calculated as ${\rm d}N/{\rm d}S=N/(\Omega\,\Delta S)$, where $N$ is the number of catalogue sources in the bin, $\Omega$ is the survey area in steradians, and $\Delta S$ is the bin width in Jy. The representative flux density of each bin was taken to be the geometric mean of its lower and upper boundaries. We present the counts in Euclidean-normalised form, $S^{2.5}{\rm d}N/{\rm d}S$, in units of Jy$^{1.5}$\,sr$^{-1}$.

The raw counts were corrected using the effective source-count correction factors derived from the source-injection simulations in Section~\ref{sec:completeness}. These corrections follow the input--output approach of \citet{Smolcic2017a} in order to account for sources that are not recovered and for the flux-density redistribution between bins. In addition, the final corrections incorporate the morphology-dependent source-recovery response derived from the extended-source simulations. For each flux-density bin, the corrected source number is $N_{\rm corr}=N_{\rm raw}f_{\rm corr}$, where $f_{\rm corr}=C_{\rm eff}^{-1}=N_{\rm in}/N_{\rm out}$.

Here $N_{\rm out}$ represents the morphology-weighted recovered population binned according to the adopted measured-flux prescription. Because the correction includes migration between flux-density bins, $f_{\rm corr}$ can be smaller than unity. Poisson uncertainties on the observed source numbers were propagated through the correction factors. The uncertainty on $f_{\rm corr}$, including both simulation-realisation and morphology-model contributions, was propagated separately and combined in quadrature with the Poisson contribution to obtain the uncertainties on the corrected counts.

Figure~\ref{fig:sourcecounts} presents the raw and corrected Euclidean-normalised source counts, which are also listed in Table~\ref{tab:source_counts}. The shaded region shows the contribution from the correction-factor uncertainty. Below 0.398\,mJy, the recovery fraction is too small to derive a robust finite correction, and we therefore report only the raw counts. The first corrected bin (0.398--0.736\,mJy) lies close to the detection limit and should be interpreted with caution. For $S>100$\,mJy, beyond the flux-density range covered by the simulations, we assume that source-recovery and flux-redistribution effects are negligible and adopt $f_{\rm corr}=1$.

We adopt the LoTSS DR3 source counts  \citep{Shimwell2026} as our primary low-frequency comparison because of their similar observing frequency and the substantially larger survey area, which provides high source statistics and minimises the impact of field-to-field variations. The LoTSS DR3 comparison sample combines counts derived from individual-pointing simulations below 0.343\,mJy, mosaic simulations between 0.343 and 3.449\,mJy, and the observed raw counts above 3.449\,mJy. The quoted LoTSS uncertainties combine the uncertainty components provided with the published DR3 source counts.

We also compare the 144 MHz source counts with the VLA-COSMOS 3\,GHz source counts from \citet{Smolcic2017a}. These were shifted to 144\,MHz assuming a power-law radio spectrum, $S_{\nu}\propto\nu^{-\alpha}$, with $\alpha=0.7$. Both the flux-density scale and the Euclidean-normalised counts were transformed consistently as $S_{144}=S_{3\,{\rm GHz}}(144/3000)^{-\alpha}$ and $(S^{2.5}{\rm d}N/{\rm d}S)_{144}=(S^{2.5}{\rm d}N/{\rm d}S)_{3\,{\rm GHz}}(144/3000)^{-1.5\alpha}$.

The corrected COSMOS 144\,MHz counts are in good agreement with the LoTSS DR3 within the uncertainties over the flux-density range where the source-count corrections are well constrained, and are also consistent with the scaled VLA-COSMOS 3 GHz distribution. The simulation-based correction substantially increases the counts close to the detection limit, while above approximately 1--2\,mJy the corrections are generally modest and the raw and corrected counts converge. The remaining bin-to-bin differences with the comparison surveys are not unexpected given the relatively small 2.6\,deg$^{2}$ area of the COSMOS field, the resulting source statistics, differences in angular resolution and source extraction, and the spectral-index assumption required to scale the 3\,GHz measurements to 144\,MHz. At the bright end, the small number of sources per bin leads to correspondingly large Poisson uncertainties.

A comparison between the measured 144\,MHz flux densities and those predicted from matched 3\,GHz sources gives a median ratio of $S_{144}^{\rm meas}/S_{144}^{\rm pred}=0.9263^{+0.0132}_{-0.0132}$ (see Figure~\ref{fig:flux_scale_test} in Appendix~\ref{app:fluxscale}), where the uncertainty on the median was estimated by bootstrap resampling. Applying the more conservative ${\rm S/N}\geq5$ threshold gives a consistent median ratio of $1.05\pm0.04$. The comparison is therefore consistent with the adopted LOFAR flux-density scale, within the limitations of the assumed spectral scaling. We additionally investigated the potential effect of source blending arising from the lower angular resolution of the LOFAR observations. Using an elliptical $15\arcsec\times6\arcsec$ search region matched to the LOFAR restoring beam, 102 of the 1984 LOFAR sources (5.1\%) contain more than one VLA-COSMOS 3\,GHz source (see Table~\ref{tab:match_stats_all}). A detailed deblending method using the higher-resolution radio data is presented in Doukas et al. (submitted), where it is applied to construct and analyse the radio spectral energy distributions of the COSMOS radio-source population. This provides an estimate of the fraction of LOFAR detections potentially affected by blending at the resolution of the 144\,MHz image. The multiplicity fraction remains small across the faint flux-density bins, indicating that blending may affect individual sources but is unlikely to be the dominant driver of the observed source-count behaviour. The flux-density-scale and blending tests are presented in detail in Appendix~\ref{app:fluxscale}. 

\begin{table}
\centering
\renewcommand{\arraystretch}{1.18}
\caption{Euclidean-normalised differential source counts at 144\,MHz. Raw counts are shown for all flux-density bins, while corrected counts are reported where a finite source-count correction is available. For $S>100$\,mJy, the catalogue is assumed to be complete and $f_{\rm corr}=1$ is adopted. The first bin is not completeness corrected but is also presented because the corresponding measurement is  shown in Figure~\ref{fig:sourcecounts}.}
\label{tab:source_counts}
\hspace*{-2mm}%
\begin{tabular}{ccccc}
\hline\hline
Flux density & Raw counts & Corrected counts & $N$ & $f_{\rm corr}$ \\
(mJy) & \multicolumn{2}{c}{(Jy$^{1.5}$\,sr$^{-1}$)} &  &  \\
\hline
0.293 & $4.6\pm0.2$ & \ldots & 454 & \ldots \\
0.541 & $12.5\pm0.6$ & $35^{+8}_{-2}$ & 491 & 2.786 \\
1 & $24\pm1$ & $45^{+3}_{-3}$ & 367 & 1.904 \\
1.85 & $52\pm3$ & $55^{+4}_{-4}$ & 321 & 1.074 \\
3.41 & $56\pm5$ & $51^{+5}_{-5}$ & 138 & 0.919 \\
6.31 & $79\pm9$ & $77^{+9}_{-9}$ & 78 & 0.976 \\
11.7 & $102\pm16$ & $113^{+18}_{-18}$ & 40 & 1.106 \\
21.5 & $186\pm34$ & $193^{+36}_{-36}$ & 29 & 1.042 \\
39.8 & $354\pm75$ & $366^{+83}_{-82}$ & 22 & 1.034 \\
73.6 & $485\pm140$ & $515\pm152$ & 12 & 1.062 \\
141 & $760\pm269$ & $760\pm269$ & 8 & 1.000 \\
316 & $2603\pm785$ & $2603\pm785$ & 11 & 1.000 \\
866 & $2644\pm1526$ & $2644\pm1526$ & 3 & 1.000 \\
\hline
\end{tabular}
\renewcommand{\arraystretch}{1.0}
\end{table}

\section{Conclusions and Outlook}
\label{sec:conclusion}

We have presented the first 144\,MHz LOFAR observations of the COSMOS field obtained as part of the DDT19\_002 programme. The observations comprise twelve 4-hour blocks obtained with the Dutch LOFAR array and processed using the standard LoTSS calibration and imaging pipeline. The final 48-h image reaches a median sensitivity of $\sim100~\mu$Jy~beam$^{-1}$ at an angular resolution of $15''\times6''$, making it one of the deepest low-frequency radio observations of the COSMOS field to date.

To assess the completeness of the resulting source catalogue extracted with {\sc PyBDSF}, we performed simulations sampling a broader range of intrinsic source sizes, accounting for the dependence of extended-source recovery on angular extent and surface brightness. These simulations were used to characterise the {\sc PyBDSF} response as a function of intrinsic angular size and were weighted to reproduce the observed $S_{\rm T}/S_{\rm P}$ distributions and resolved-source fractions of the COSMOS population. The resulting morphology-weighted extended-source population replaces the corresponding extended component of the baseline simulations and was used to derive the final source-count corrections. Tests using the observed broadening of apparently compact sources indicate that residual calibration or ionospheric broadening can affect individual resolved/unresolved classifications, while the derived source counts change by less than 5\% in most bins and by at most $\sim18\%$ in the most affected bin. We find 80\% and 90\% completeness at approximately 1.13 and 1.48\,mJy, respectively, with the completeness approaching unity above $\sim2$--$3$\,mJy. Below 0.398\,mJy the recovery fraction is too small to derive a robust finite source-count correction. The first corrected bin, at 0.398--0.736\,mJy, remains particularly sensitive to the morphology-dependent correction. At the ${\rm S/N}\geq5$ level adopted for the  catalogue analysis, the negative-image test gives a catalogue reliability of 98.5\%, increasing to $99.6\%$ at ${\rm S/N}\ge8$.

The Euclidean-normalised source counts are in good agreement with LoTSS DR3 within the uncertainties over the flux-density range where the source-count corrections are well constrained, and are also consistent with the VLA-COSMOS 3\,GHz counts scaled to 144\,MHz. The remaining differences may arise from limited source statistics, field-to-field variations, differences in angular resolution and source extraction, and, for the 3\,GHz comparison, the assumed spectral scaling. For the 980 sources cross-matched with the VLA-COSMOS 3\,GHz catalogue within $2$\arcsec, we find a median 144\,MHz--3\,GHz spectral index of $\alpha_{144\,\mathrm{MHz}}^{3\,\mathrm{GHz}}=0.67$, consistent with the value of 0.7 adopted to scale the 3\,GHz source counts. The spectral index shows no significant dependence on signal-to-noise ratio or distance from the pointing centre. An apparent flattening towards faint 144\,MHz flux densities is also observed, although the decreasing 3\,GHz counterpart fraction at the faint end means that this trend is affected by the 3\,GHz selection and cannot be interpreted as an intrinsic spectral-index--flux-density relation. The astrometric comparison with the VLA-COSMOS 3\,GHz catalogue shows no significant field-dependent trends. The measured median offsets are $\Delta{\rm RA}=+0.35$\arcsec\ and $\Delta{\rm Dec}=-0.83$\arcsec, corresponding to approximately 6\% of the respective synthesised-beam dimensions. Astrometrically corrected coordinates relative to the VLA-COSMOS 3\,GHz reference frame are provided in the released catalogue, while the original LOFAR positions are retained in separate columns. We also find no evidence for a substantial systematic offset in the LOFAR flux-density scale. Approximately 5\% of the LOFAR sources contain multiple VLA-COSMOS 3\,GHz counterparts within a beam-matched search region; consequently, the simple cross-matched 3\,GHz flux densities and spectral indices provided in the catalogue should be interpreted with caution for blended or complex sources. The catalogue presented here adds the first LOFAR 144\,MHz source catalogue of COSMOS to the extensive radio and multiwavelength coverage of the field. It can be used to study the faint radio-source population and provides a reference for future, deeper low-frequency observations of COSMOS.

\subsection{Outlook}

The observations presented in this paper represent the first stage of a broader low-frequency legacy programme for the COSMOS field. Despite being among the deepest LOFAR observations currently available, the present 48-hour dataset remains sensitivity-limited compared with the wealth of multiwavelength observations in COSMOS. Many of the faint star-forming galaxies and AGN detected at GHz frequencies remain beyond the reach of the current observations, while the low-frequency emission accessible with LOFAR provides a complementary view of synchrotron-emitting particle populations and magnetic fields.

Substantially deeper observations with LOFAR2.0 would extend this work to much fainter radio-source populations. The proposed (2023) ECOLE (Exploring COSMOS with LOFAR's Eyes) LOFAR2.0 Large Programme\footnote{The programme decision remains pending at the time of writing.} illustrates the potential of future deeper LOFAR2.0 observations of COSMOS. The sensitivity of the present observations in the context of existing and proposed radio-continuum observations of COSMOS is shown in Figure~\ref{fig:future_surveys} in Appendix~\ref{app:radio_surveys}.

The catalogue presented here provides the foundation for future studies combining the LOFAR observations with the extensive ancillary data available across COSMOS, enabling a plethora of AGN and star-forming galaxy studies. Forthcoming work will add deblending, multiwavelength identifications, redshifts, source assembly, and physical classifications, while processing of the international LOFAR baselines will provide sub-arcsecond imaging for improved source identification and morphological characterisation.

\begin{acknowledgements} 

We thank Aleksander Kurek and Dominik J. Bomans for helpful comments and suggestions on an earlier version of the manuscript.  MJH acknowledges support from  STFC [ST/Y001249/1]. LKM is grateful for support from a UKRI FLF [MR/Y020405/1] and LOFAR-UK via STFC [ST/V002406/1]. VS acknowledges support by the European Regional Development Fund under grant agreement PK.1.1.10.0007 (DATACROSS). CLH acknowledges support from the Science and Technology Facilities Council (STFC) through grant ST/Y000951/1. MV acknowledges financial support from the Inter-University Institute for Data Intensive Astronomy (IDIA), a partnership of the University of Cape Town, the University of Pretoria and the University of the Western Cape, and from the South African Department of Science and Innovation's National Research Foundation under the ISARP RADIOMAP Joint Research Scheme (DSI-NRF Grant Number 150551) and the CPRR HIPPO Project (DSI-NRF Grant Number SRUG22031677). 
MA is supported by FONDECYT grant number 1252054, and gratefully acknowledges support from ANID Basal Project FB210003,  ANID MILENIO NCN2024\_112 and ANID + Vinculaci\'on Internacional + FOVI250261. ET acknowledges support from the ANID CATA-BASAL program FB210003, and FONDECYT Regular 1241005 and  1250821. LOFAR is the Low Frequency Array designed and constructed by ASTRON and is collectively operated by the LOFAR ERIC under a joint scientific policy. The use of AI tools was limited to language editing assistance. 

\end{acknowledgements}

 \bibliographystyle{aa} 
   \bibliography{refs}

\appendix

\section{Catalogue robustness tests}
\label{app:completeness_tests}

We tested the sensitivity of the completeness to the positional matching radius used to associate injected and recovered sources. The fiducial analysis presented in Section~\ref{sec:completeness} adopts a matching radius of $3^{\prime\prime}$. To assess whether this choice affects the inferred completeness, we repeated the full matching procedure using radii of $2^{\prime\prime}$ and $4^{\prime\prime}$, while keeping the injected source population, recovered source catalogues, peak ${\rm S/N}\geq5$ recovery criterion, and flux-density bins unchanged.

\begin{figure*}
    \centering
    \includegraphics[width=\linewidth]{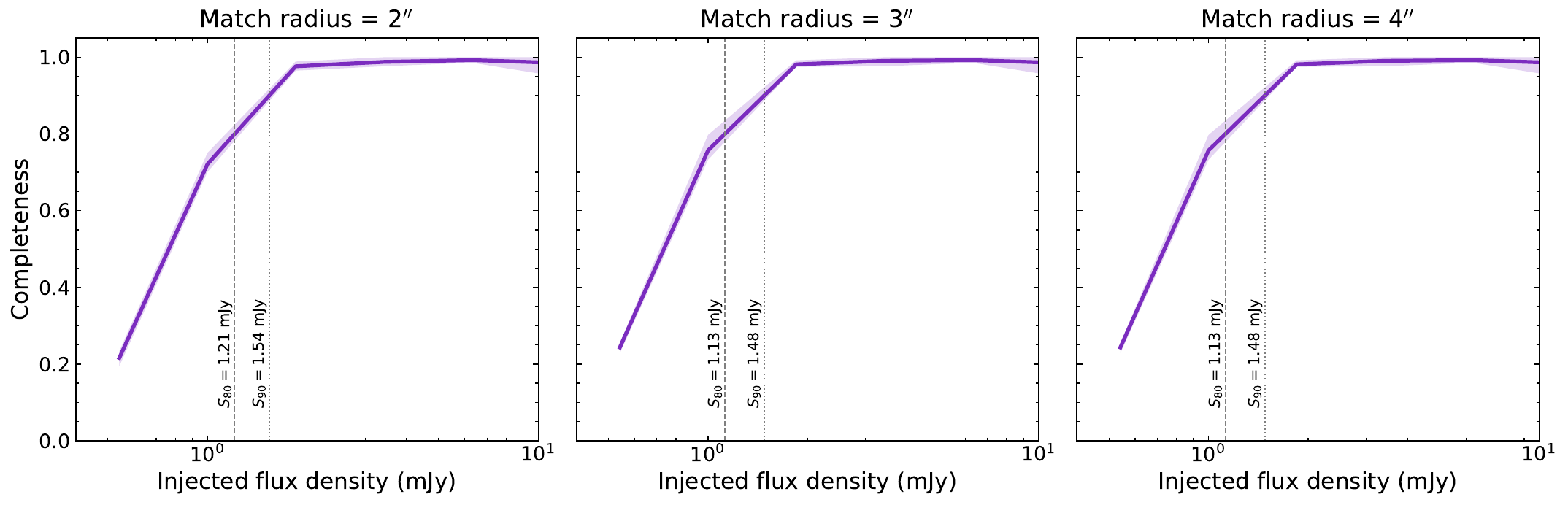}
    \caption{Completeness as a function of injected flux density for positional matching radii of $2^{\prime\prime}$ (left), $3^{\prime\prime}$ (centre), and $4^{\prime\prime}$ (right). In each panel, the solid purple curve shows the completeness calculated from the combined injected and recovered     populations across the ten simulation realisations, while the shaded region indicates the $16^{\rm th}$--$84^{\rm th}$ percentile range among the individual realisations. Vertical lines mark the flux densities corresponding to 80\% and 90\% completeness. The $3^{\prime\prime}$ and     $4^{\prime\prime}$ results are effectively identical, while the more restrictive $2^{\prime\prime}$ criterion produces a modest decrease in completeness close to the detection threshold. The comparison supports the adoption of $3^{\prime\prime}$ as the fiducial matching radius.
}
\label{fig:completeness_matching}
\end{figure*}

Figure~\ref{fig:completeness_matching} compares the completeness obtained for the three matching radii. For each flux-density bin, the completeness is defined as $C_{\rm rec} = \frac{N_{\rm rec}}{N_{\rm in}}$, where $N_{\rm in}$ is the number of injected sources and $N_{\rm rec}$ is the number associated with a recovered source within the specified matching radius and satisfying the recovered peak ${\rm S/N}\geq5$ criterion. The central curves are calculated from the combined injected and recovered populations across the ten simulation realisations, while the shaded regions show the $16^{\rm th}$--$84^{\rm th}$ percentile range among the individual realisations.

The completeness curves obtained with matching radii of $3^{\prime\prime}$ and $4^{\prime\prime}$ are effectively identical over the full flux-density range. Adopting the more restrictive $2^{\prime\prime}$ radius produces a modest reduction in completeness close to the detection threshold, where positional uncertainties are largest. The differences rapidly decrease with increasing flux density, and all three matching criteria converge in the regime where the catalogue approaches full recovery.

For the fiducial $3^{\prime\prime}$ matching radius, the catalogue reaches 80\% and 90\% completeness at approximately 1.13\,mJy and 1.48\,mJy, respectively. The close agreement between the $3^{\prime\prime}$ and $4^{\prime\prime}$ results indicates that increasing the matching radius does not recover an appreciable additional population of ${\rm S/N}\geq5$ sources. Conversely, the slightly lower completeness obtained with $2^{\prime\prime}$ indicates that this more restrictive criterion excludes a small number of otherwise valid associations, particularly at low flux densities. We therefore retain $3^{\prime\prime}$ as the fiducial matching radius.

This test demonstrates that the completeness adopted in Section~\ref{sec:completeness} is insensitive to reasonable variations in the positional association criterion. The strong decline in $C_{\rm rec}$ towards faint flux densities is therefore intrinsic to the source-recovery statistics of the simulations rather than being driven by the adopted $3^{\prime\prime}$ matching radius.

As a complementary test of the catalogue robustness, we also assessed the false-detection rate using the inverted-image analysis described in Section~\ref{sec:false_detections}. The negative-image extraction was performed using the same rms maps derived from the positive science image and supplied to the catalogue extraction, ensuring a consistent local noise treatment. Table~\ref{tab:false_detection} gives the differential and cumulative false-detection fractions and the corresponding cumulative catalogue reliability as a function of peak S/N. At the ${\rm S/N}\geq5$ threshold adopted for the simulation-based resolved/unresolved classification and source-count analysis, we find 14 negative detections compared with 906 positive sources, corresponding to a reliability of 98.5\%. The reliability exceeds 99\% at ${\rm S/N}\geq5.5$ and reaches 99.6\% at ${\rm S/N}\geq8$.

\begin{table}
\centering
\renewcommand{\arraystretch}{1.18}
\caption{Differential and cumulative false-detection fractions and catalogue reliabilities derived from the negative-image test. The negative-image extraction uses the same rms maps derived from the positive science image as the real catalogue extraction. $N_{\rm pos}$ and $N_{\rm neg}$ are the cumulative numbers of positive- and negative-image detections above each peak signal-to-noise threshold. Differential quantities refer to the interval between each listed threshold and the next, with the final interval corresponding to ${\rm S/N}\geq10$.
}
\label{tab:false_detection}
\setlength{\tabcolsep}{4pt}
\begin{tabular}{cccccc}
\hline
S/N threshold & $N_{\rm pos}$ & $N_{\rm neg}$ & $f_{\rm false}^{\rm diff}$ (\%) & $f_{\rm false}^{\rm cum}$ (\%) & $R^{\rm cum}$ (\%) \\
\hline
$\geq 4.0$ & 1245 & 51 & 14.4 & 4.1 & 95.9 \\
$\geq 4.5$ & 1064 & 25 & 7.0 & 2.3 & 97.7 \\
$\geq 5.0$ & 906 & 14 & 7.5 & 1.5 & 98.5 \\
$\geq 5.5$ & 826 & 8 & 4.8 & 1.0 & 99.0 \\
$\geq 6.0$ & 764 & 5 & 2.0 & 0.7 & 99.3 \\
$\geq 7.0$ & 666 & 3 & 1.1 & 0.5 & 99.5 \\
$\geq 8.0$ & 571 & 2 & 0.9 & 0.4 & 99.6 \\
$\geq 10.0$ & 460 & 1 & 0.2 & 0.2 & 99.8 \\
\hline
\end{tabular}
\end{table}

\section{Additional catalogue validation tests}
\label{app:cat_tests}

\subsection{Flux-density scale and source blending}
\label{app:fluxscale}

As an additional consistency check, we compared the measured LOFAR 144\,MHz flux densities with those predicted from the VLA-COSMOS 3\,GHz flux densities for the same 980 sources matched within $2$\arcsec\ described in Section~\ref{sec:spectral_indices}. The predicted 144\,MHz flux densities depend directly on the assumed spectral index. We adopt $\alpha=0.7$, consistent with the median $\alpha_{144\,\mathrm{MHz}}^{3\,\mathrm{GHz}}=0.67$ measured for this cross-matched sample. This comparison is not intended as an independent absolute flux-scale calibration, since individual sources exhibit a range of spectral indices and may also be affected by spectral curvature, variability, and differences in angular resolution between the two surveys. 

\begin{figure}[h!]
    \centering
    \resizebox{\hsize}{!}
 {
\includegraphics[width=\linewidth]{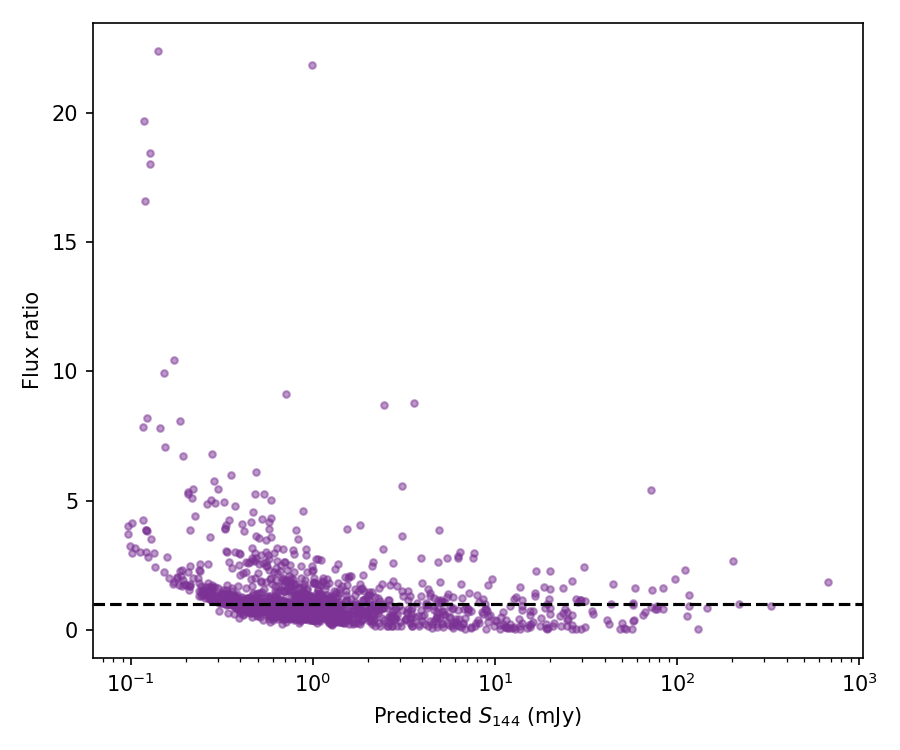}}
 \resizebox{\hsize}{!}
 {
\includegraphics[width=\linewidth]{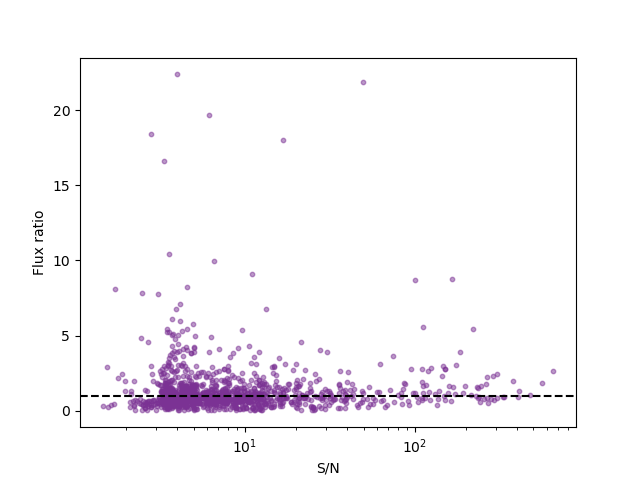}}  
\caption{
Comparison of the measured LOFAR 144\,MHz flux densities with those predicted from the matched VLA-COSMOS 3\,GHz catalogue assuming a power-law radio spectrum with $\alpha=0.7$. At 144\,MHz, we adopt the peak flux density for unresolved sources and the integrated flux density for resolved sources. The upper panel compares the measured and predicted flux densities, while the lower panel shows their ratio as a function of LOFAR peak signal-to-noise ratio. The median flux density ratio, $S_{144}^{\rm meas}/S_{144}^{\rm pred}=0.9263^{+0.0132}_{-0.0132}$, is close to unity and shows no evidence for a substantial systematic offset in the LOFAR flux-density scale, within the limitations of the assumed spectral index.
}
\label{fig:flux_scale_test}
\end{figure}

Figure~\ref{fig:flux_scale_test} compares the measured and predicted 144\,MHz flux densities and shows their ratio as a function of signal-to-noise ratio. At 144\,MHz, we adopt the peak flux density for unresolved sources and the integrated flux density for resolved sources. For the adopted $\alpha=0.7$, the distribution is centred close to unity, with a median ratio of $S_{144}^{\rm meas}/S_{144}^{\rm pred}=0.9263^{+0.0132}_{-0.0132}$. The uncertainty on the median was estimated by bootstrap resampling. The scatter about the median is expected given the diverse radio-source population, with sources exhibiting a range of spectral indices, as well as the different angular resolutions of the LOFAR and VLA observations. The absence of a systematic trend in the flux-density ratio towards low S/N also provides no evidence for a sensitivity-dependent bias in the LOFAR flux-density measurements.

A tail towards $S_{144}^{\rm meas}/S_{144}^{\rm pred}>1$ is present. This may partly reflect the range of intrinsic spectral indices, while another possible contribution is source blending. The VLA-COSMOS 3\,GHz observations have substantially higher angular resolution than the $15\arcsec\times6\arcsec$ LOFAR image, and multiple 3\,GHz sources may therefore be detected as a single LOFAR component. In such cases, the integrated 144\,MHz flux density can contain emission from more than one 3\,GHz source. Blending can consequently both increase the measured LOFAR flux density relative to that predicted from an individual counterpart and reduce the number of components entering the 144\,MHz source counts.

To quantify the importance of this effect, we counted the number of VLA-COSMOS 3\,GHz sources surrounding each of the 1984 LOFAR sources using several search regions. We considered circular radii of $6\arcsec$ and of $15\arcsec$ to probe nearby source multiplicity on progressively larger angular scales. Additionally, we adopt an elliptical $15\arcsec\times6\arcsec$ region matched to the FWHM footprint of the LOFAR synthesised beam. The results are summarised in Table~\ref{tab:match_stats_all}. Within a circular radius of $6\arcsec$, 140 LOFAR sources ($7.1\%$) have more than one 3\,GHz source, increasing to 420 ($21.2\%$) within $15\arcsec$. Using the beam-matched elliptical region, 102 sources ($5.1\%$) contain multiple 3\,GHz counterparts. The elliptical region provides the closest approximation to the area over which blending is expected to affect a LOFAR detection, while the circular searches illustrate the dependence of the multiplicity fraction on the adopted matching geometry.

We additionally examined the multiplicity fraction as a function of 144\,MHz flux density using the beam-matched $15\arcsec\times6\arcsec$ elliptical region adopted in Table~\ref{tab:match_stats_all}. The fraction of sources with multiple 3\,GHz counterparts remains small across the faint flux-density bins, increasing from $3.7\%$ at $0.22$--$0.40$\,mJy to $6.9\%$ at $1.36$--$2.51$\,mJy. Even if all additional 3\,GHz counterparts are assumed to represent sources blended into a single LOFAR detection, the corresponding increase in the underlying source numbers is only $\sim4$--$8\%$. Blending can therefore affect the flux densities and multiplicities of individual sources and introduce a bias in the measured source counts. However, the relatively small fraction of sources with multiple 3\,GHz counterparts suggests that blending is unlikely to be the dominant driver of the overall source-count behaviour.

\begin{table*}
\centering
\caption{Summary of the 144\,MHz--3\,GHz counterpart statistics for different search regions. For each LOFAR source, the number of VLA-COSMOS 3\,GHz sources within the specified search region was counted.} 
\label{tab:match_stats_all}
\begin{tabular}{lccccc}
\hline\hline
Search region & Total LOFAR & 0 counterparts & 1 counterpart & $>1$ counterparts & Fraction with $>1$ \\
 & sources checked &  &  &  & counterparts (\%) \\
\hline
Circular, $6\arcsec$ radius & 1984 & 559 & 1285 & 140 & 7.1 \\
Circular, $15\arcsec$ radius & 1984 & 437 & 1127 & 420 & 21.2 \\
Elliptical, $15\arcsec\times6\arcsec$ & 1984 & 642 & 1240 & 102 & 5.1 \\
\hline
\end{tabular}
\end{table*}

\begin{figure}
    \centering
    \includegraphics[width=\linewidth]{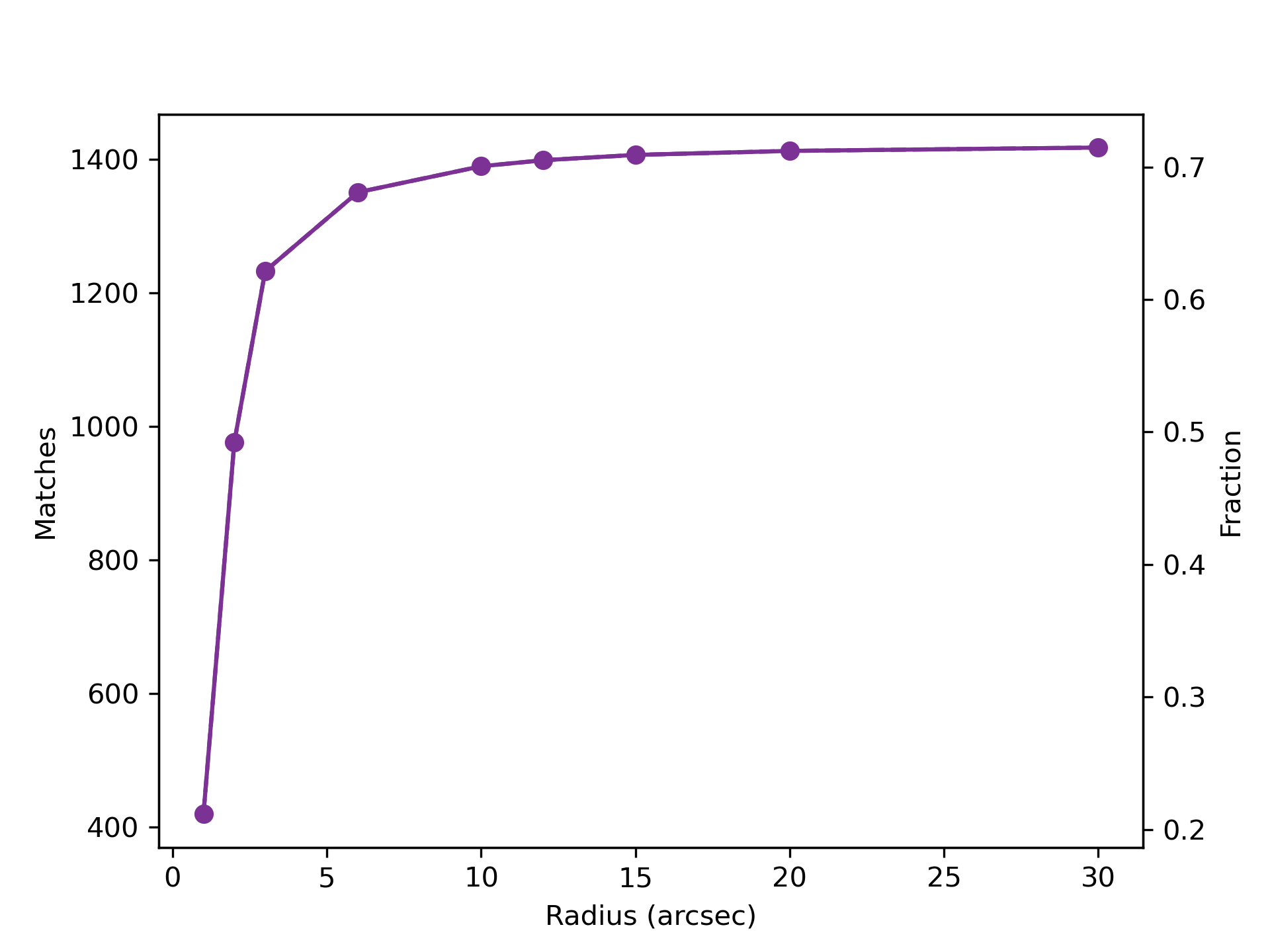}
    \caption{
Dependence of the LOFAR--VLA-COSMOS cross-matching statistics on the adopted search radius. The comparison illustrates how the number of candidate 3\,GHz counterparts associated with a LOFAR source changes as progressively larger regions around the 144\,MHz position are considered. 
}
\label{fig:matching_radius}
\end{figure}

Figure~\ref{fig:matching_radius} provides a complementary view of the cross-matching behaviour as a function of separation between the LOFAR and VLA-COSMOS positions. For each search radius, we count the number of sources in the full 1984-source LOFAR catalogue with at least one VLA-COSMOS 3\,GHz source within that radius; the corresponding fraction is calculated relative to the full LOFAR catalogue. The majority of associations occur at small separations, while progressively larger search regions include an increasing number of additional candidate counterparts. This illustrates the trade-off between recovering genuine associations affected by positional uncertainties or differences in source morphology and including unrelated neighbouring 3\,GHz sources. We emphasise that these statistics describe positional associations rather than a complete counterpart-identification analysis, which will be presented in forthcoming work.

These tests provide no evidence for a substantial systematic offset in the LOFAR flux-density scale, within the limitations of the assumed spectral scaling. The median ratio between the measured and predicted 144\,MHz flux densities is close to unity and shows no significant dependence on S/N. Source blending is present and can affect individual flux-density measurements and source multiplicities, but the observed counterpart statistics indicate that its overall contribution to the source counts is relatively modest.

The source-injection analysis presented in Section~\ref{sec:completeness} explicitly considers the additional effect of source morphology on recovery. Extended sources have lower peak surface brightness than compact sources of the same integrated flux density and are therefore more susceptible to the peak S/N detection threshold. This response was characterised using the dedicated extended-source simulations and incorporated into the final morphology-weighted source-count corrections. The flux-density-scale and blending tests presented here therefore provide complementary checks on effects that are not described by the completeness correction alone.

\subsection{Choice of functional form for the $S_{\rm T}/S_{\rm P}$ envelope}
\label{app:envelope_function}

\begin{figure}
    \centering
    \includegraphics[width=\columnwidth]
    {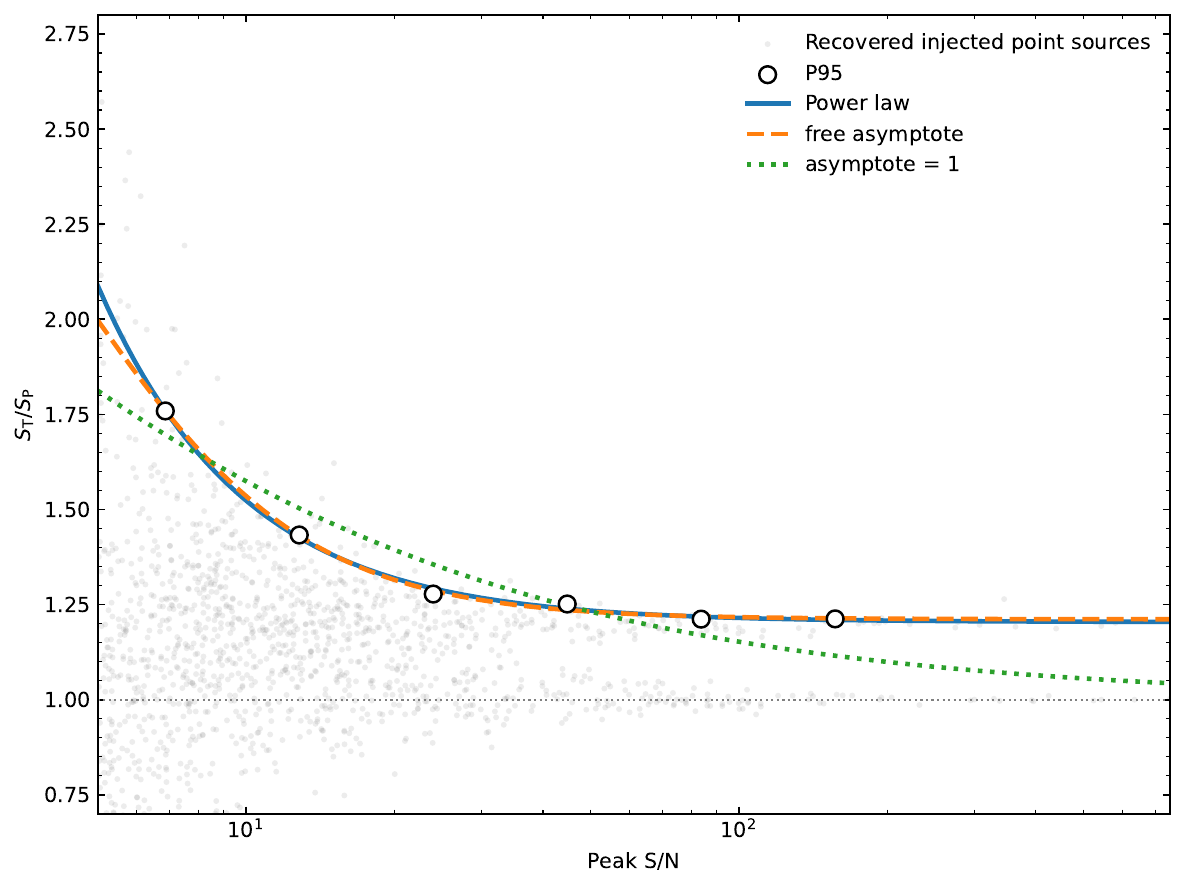}
    \caption{Comparison of alternative functional forms for the $S_{\rm T}/S_{\rm P}$ envelope. Grey points show the recovered injected point sources and open circles indicate the 95th percentile of $S_{\rm T}/S_{\rm P}$ in bins of peak signal-to-noise ratio. The solid line shows the adopted power-law parametrisation, $S_{\rm T}/S_{\rm P}  = c+a({\rm S/N})^{b}$. The dashed line shows a model with a free high-S/N asymptote, while the dotted line shows the same functional form constrained to approach $S_{\rm T}/S_{\rm P}=1$. The free-asymptote models provide very similar descriptions of the simulated point-source population and approach comparable high-S/N asymptotes of $S_{\rm T}/S_{\rm P}\simeq1.19$--$1.21$.  In contrast, forcing the envelope to approach unity provides a poorer description of the simulated percentiles and increasingly places high-S/N injected point sources above the envelope.
    }
    \label{fig:envelope_model_comparison}
\end{figure}

We tested whether the non-unity high-S/N asymptote of the adopted $S_{\rm T}/S_{\rm P}$ envelope depends on the choice of functional form. In addition to the adopted power-law model, we fitted an alternative functional form that approaches a constant value at high S/N, considering both a free asymptote and an asymptote fixed to unity, to the 95th percentiles of the recovered injected point-source population  (Figure~\ref{fig:envelope_model_comparison}). The free-asymptote model converges to $S_{\rm T}/S_{\rm P}\simeq1.21$, consistent with the high-S/N asymptote of the adopted power-law model ($S_{\rm T}/S_{\rm P}\simeq1.19$), and produces a similar point-source false-positive fraction. In contrast, forcing the envelope to approach unity provides a substantially poorer description of the simulated percentiles and results in an increasing fraction of high-S/N injected point sources being classified as resolved. We therefore retain the simpler power-law parametrisation.

\subsubsection{Tests for residual source broadening}
\label{app:source_broadening}

\begin{figure}
    \centering
    \includegraphics[width=\columnwidth]
    {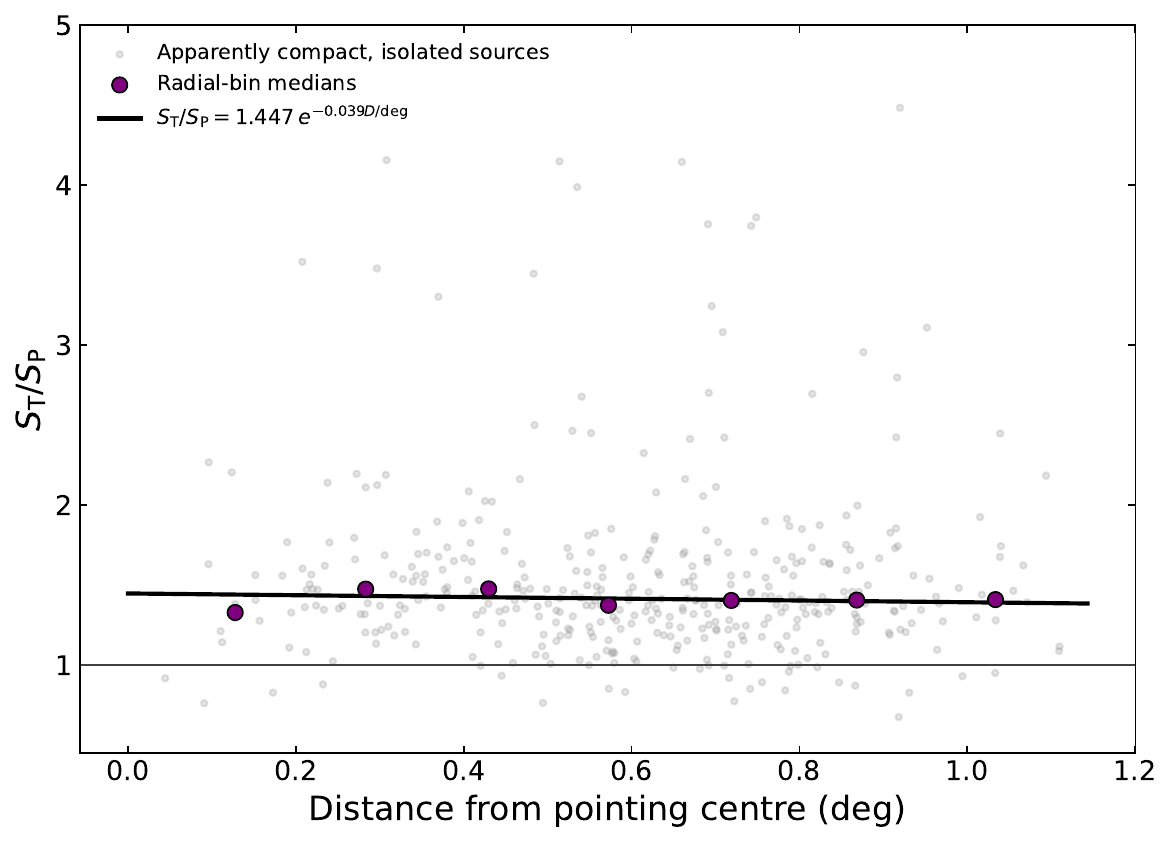}
    \caption{$S_{\rm T}/S_{\rm P}$ as a function of distance from the pointing centre for the 415 apparently compact and isolated sources. These sources are selected independently of the adopted $S_{\rm T}/S_{\rm P}$ classification, requiring a nearest-neighbour separation $>15$\arcsec, fitted major axis $<15$\arcsec, $S_{\rm T}/\sigma_{S_{\rm T}}>5$, and {\sc PyBDSF} ${\tt S\_Code}={\tt S}$. Grey points show the individual sources, purple points show the median $S_{\rm T}/S_{\rm P}$ in radial bins, and the black line shows the fitted relation $S_{\rm T}/S_{\rm P} = 1.447\,\mathrm{e}^{-0.039D}$, where $D$ is the distance from the pointing centre in degrees. The population has a median $S_{\rm T}/S_{\rm P}=1.41$, but shows no statistically significant radial dependence (Spearman $\rho=-0.07$, $p=0.18$).}
    \label{fig:compact_radial_smearing}
\end{figure}

The point-source simulations used to derive the adopted $S_{\rm T}/S_{\rm P}$ envelope reproduce the effects of the local image noise and source-fitting uncertainties, but the injected sources themselves are not subject to residual ionospheric or calibration-related broadening. We therefore examined the behaviour of an independently selected population of apparently compact and isolated catalogue sources. We require a nearest-neighbour separation $>15$\arcsec, fitted major axis $<15$\arcsec, $S_{\rm T}/\sigma_{S_{\rm T}}>5$, and {\sc PyBDSF} ${\tt S\_Code}={\tt S}$, yielding 415 sources. As shown in Figure~\ref{fig:compact_radial_smearing}, this population has a median $S_{\rm T}/S_{\rm P}=1.41$, demonstrating that real apparently compact sources are broader than the ideal injected point sources. However, we find no significant dependence of $S_{\rm T}/S_{\rm P}$ on distance from the pointing centre (Spearman $\rho=-0.07$, $p=0.18$), and therefore no empirical basis for applying a position-dependent smearing correction.

As a further test, we applied the survival-function approach used by \citet{Shimwell2025}. Our implementation is intended as a diagnostic adaptation of the \citet{Shimwell2025} procedure rather than an exact reproduction, since our simulations use ideal injected point sources rather than CLEAN-component models derived from observed apparently compact sources. Stable 5:1 survival-function thresholds were recovered in only three integrated-significance bins, at $S_{\rm T}/\sigma_{S_{\rm T}}=5.90$, 8.10, and 12.12, while no stable crossing was obtained at higher significance. The COSMOS sample therefore does not provide enough threshold points to constrain the four-parameter sigmoidal relation robustly. Using an interpolation through the measured thresholds as a diagnostic gives a resolved fraction of 57.4\% among the 610 sources satisfying the requirements of this test, compared with 58.4\% using the adopted COSMOS envelope, with 87.2\% of sources retaining the same resolved/unresolved classification.

\begin{figure}
    \centering
    \includegraphics[width=\columnwidth]
    {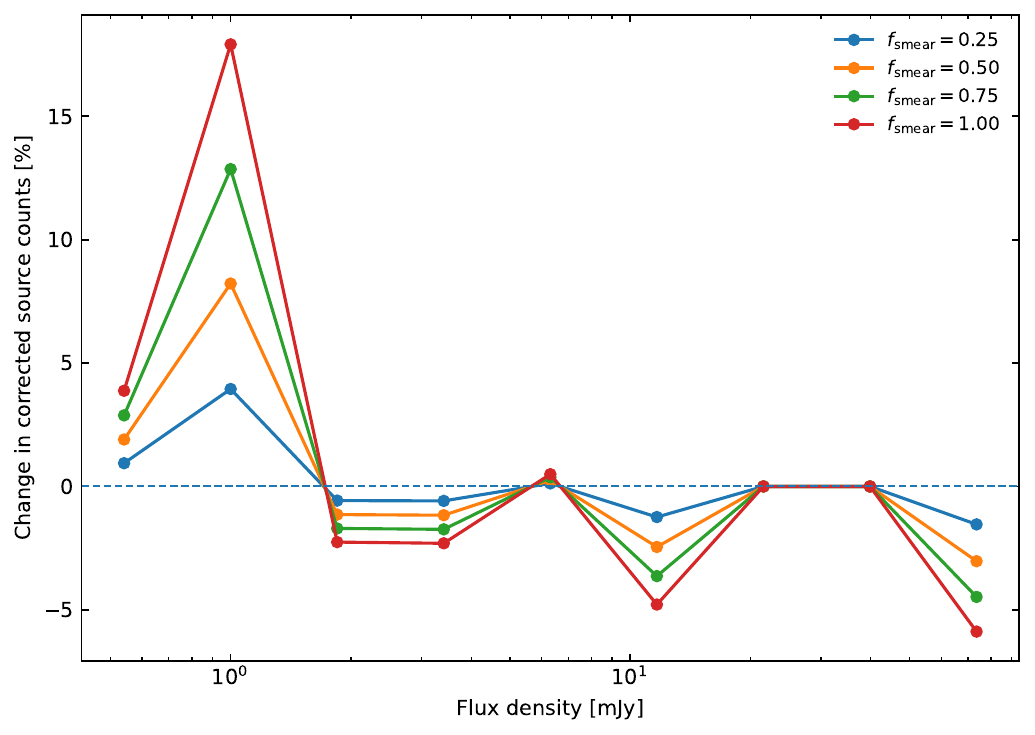}
    \caption{Effect of residual source broadening on the 144 MHz source counts. Percentage change in the corrected     Euclidean-normalised source counts relative to the fiducial morphology-informed correction ($f_{\rm smear}=0$), when fractions $f_{\rm smear}=0.25$--1.0 of the intrinsically compact simulated population are replaced by sources subject to the empirically derived broadening distribution. The effect is below 5\% in most flux-density bins, reaching approximately 18\% in the $0.736$--$1.359$\,mJy bin for the limiting $f_{\rm smear}=1$ case.}
    \label{fig:smearing_counts}
\end{figure}

\subsection{Astrometry of unresolved sources}
\label{app:astrometry_unresolved}

The astrometric analysis presented in Sect.~\ref{sec:astrometry} includes both resolved and unresolved radio sources. For extended sources, differences between the radio morphology at 144\,MHz and 3\,GHz can result in different flux-weighted source positions and therefore contribute additional scatter to the measured positional offsets. To assess this effect, we repeated the analysis using only sources classified as unresolved according to the $S_{\rm T}/S_{\rm P}$ envelope adopted in Section~\ref{sec:fluxratios}. We retain the same LOFAR--VLA matching and astrometric selection, $|\Delta{\rm RA}|\leq7.5$\arcsec\ and $|\Delta{\rm Dec}|\leq7.5$\arcsec, yielding 425 sources.

\begin{figure}
    \centering
    \includegraphics[width=\columnwidth]
    {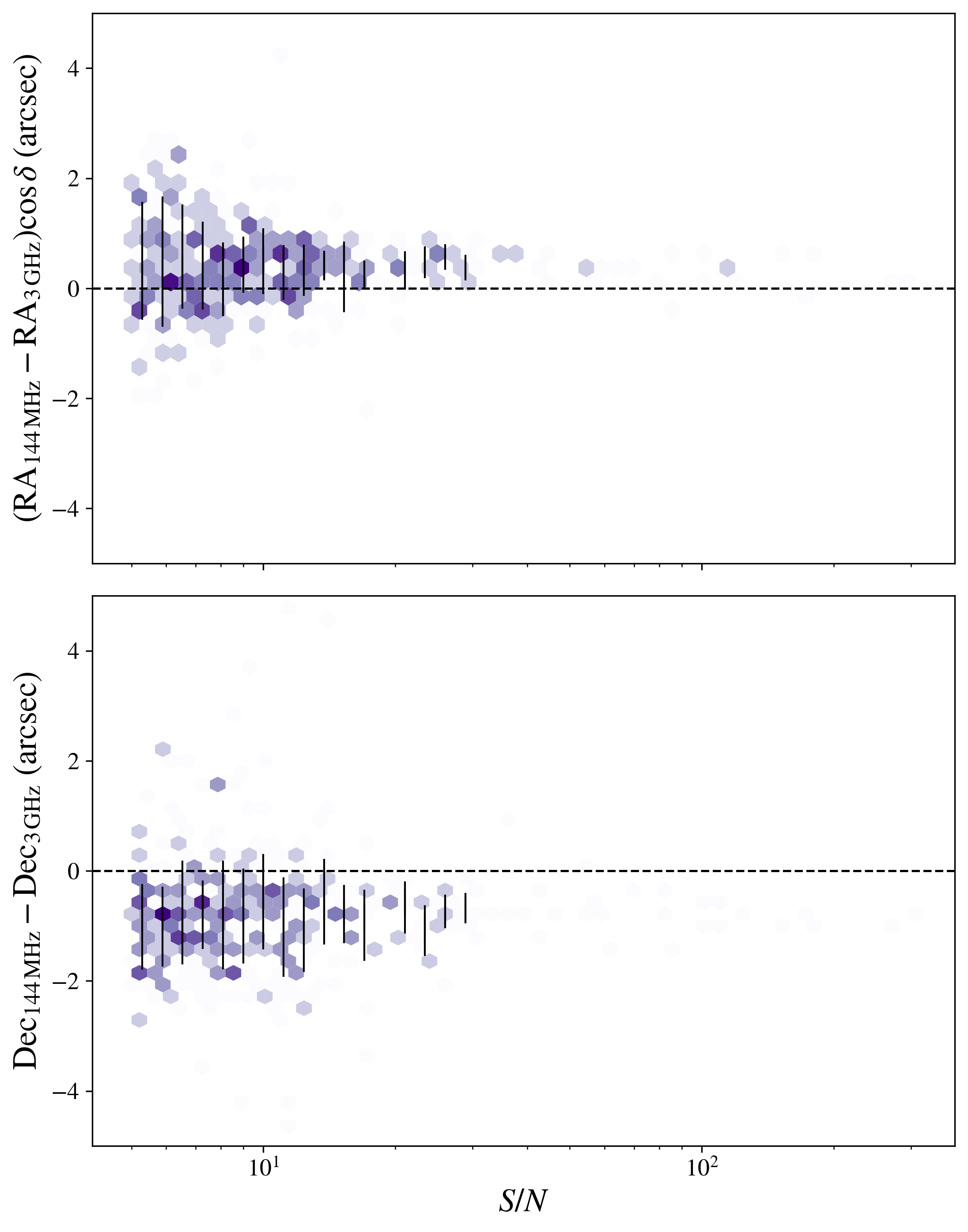}
    \caption{Astrometric offsets between the LOFAR 144\,MHz and VLA-COSMOS 3\,GHz positions as a function of LOFAR peak S/N for the 425 sources classified as unresolved according to the adopted $S_{\rm T}/S_{\rm P}$ envelope. Only sources satisfying the astrometric selection $|\Delta{\rm RA}|\leq7.5$\arcsec\ and $|\Delta{\rm Dec}|\leq7.5$\arcsec\ are shown. The colour scale indicates the source density, while black symbols show the median positional offset in S/N bins, with error bars indicating the $16^{\rm th}$--$84^{\rm th}$ percentile range. The horizontal dashed lines indicate zero positional offset.
    }
    \label{fig:astrometry_snr_unresolved}
\end{figure}

Figure~\ref{fig:astrometry_snr_unresolved} shows the positional offsets as a function of peak S/N for this subset. The median offsets are $\Delta{\rm RA}=+0.39$\arcsec\ and $\Delta{\rm Dec}=-0.78$\arcsec, similar to the values obtained for the full astrometric sample ($+0.35$\arcsec\ and $-0.83$\arcsec, respectively). The robust scatter decreases from $0.92$\arcsec\ to $0.59$\arcsec\ in RA and from $0.93$\arcsec\ to $0.73$\arcsec\ in Dec, while the radius enclosing 68\% of the sources decreases from $1.71$\arcsec\ to $1.15$\arcsec. The smaller scatter is consistent with extended-source morphology contributing to the positional dispersion, whereas the similarity of the median offsets shows that it does not drive the systematic astrometric offset.

\begin{figure}[t!]
    \centering

    \begin{minipage}[t]{0.49\textwidth}
        \centering
        \includegraphics[width=\linewidth]
        {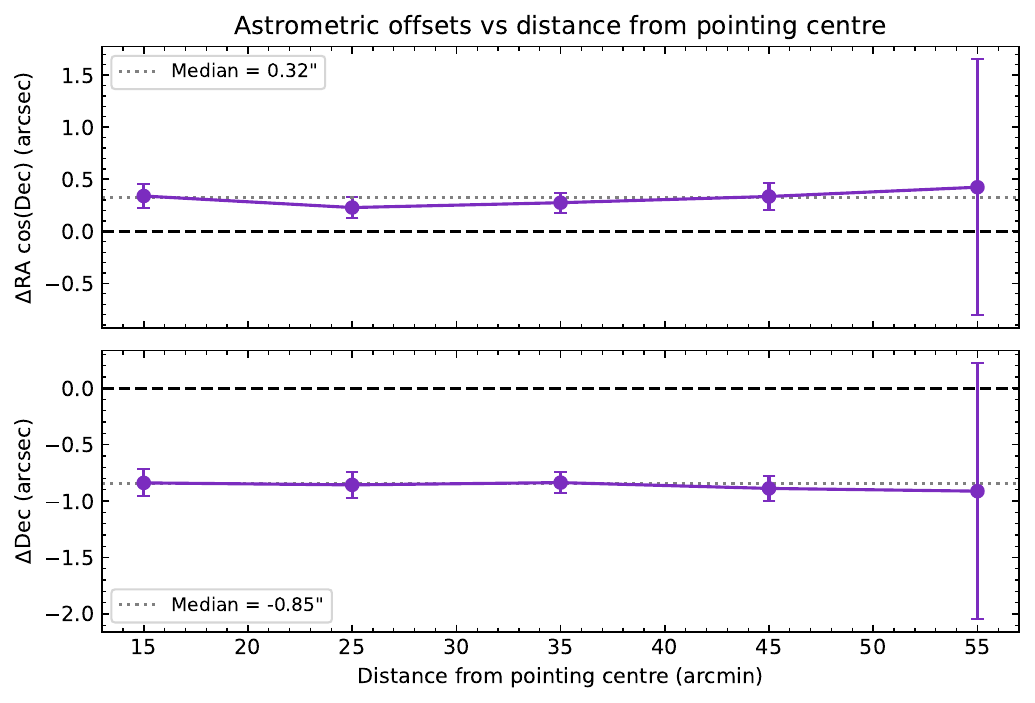}
    \end{minipage}
    \hfill
    \begin{minipage}[t]{0.49\textwidth}
        \centering
        \includegraphics[width=\linewidth]
        {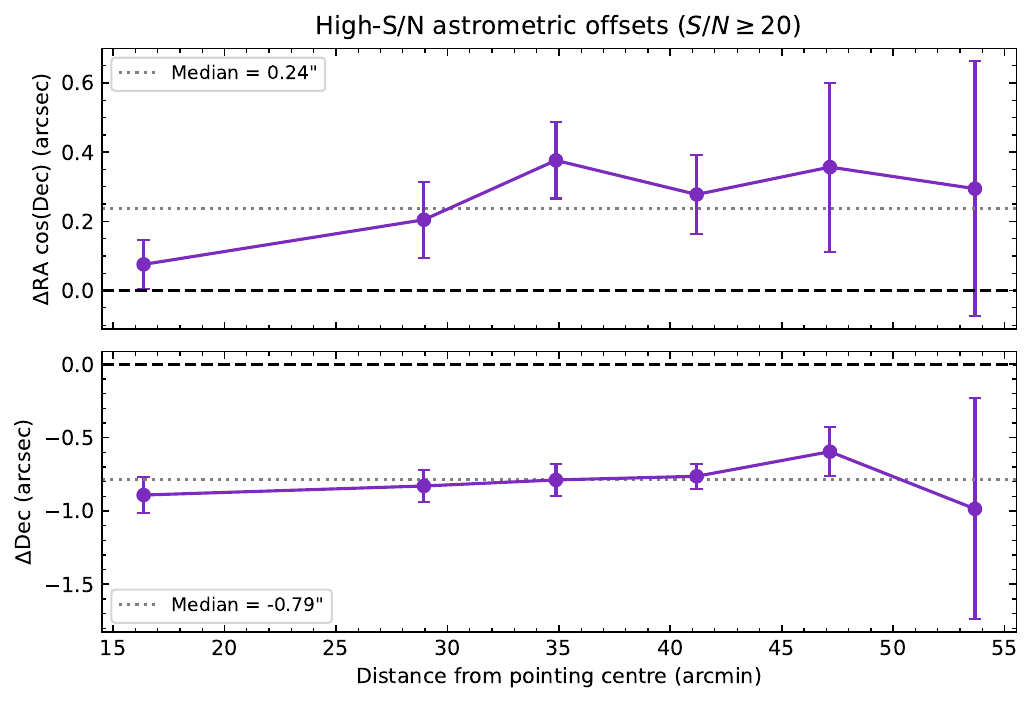}
    \end{minipage}

    \caption{Astrometric offsets as a function of distance from the LOFAR pointing centre. Top pair: full matched sample. Bottom pair: sources with $S/N\geq20$. Within each pair, the upper and lower panels show the median $\Delta{\rm RA}\cos({\rm Dec})$ and $\Delta{\rm Dec}$, respectively. Error bars indicate uncertainties on the median. The full sample uses fixed $10\arcmin$ radial bins and the high-S/N sample equal-population bins. Grey dotted and black dashed lines indicate the global median and zero offset, respectively. No significant radial dependence is observed.
    }
    \label{fig:astrometry_radius}
\end{figure}

As an additional test for spatially dependent astrometric systematics, we examined the LOFAR--VLA-COSMOS positional offsets as a function of distance from the pointing centre. Figure~\ref{fig:astrometry_radius} shows the median $\Delta$RA and $\Delta$Dec offsets in radial bins of width $10\arcmin$. The top panels show the full matched sample, while the bottom panels restrict the analysis to sources with $S/N\geq20$, for which the contribution from random positional uncertainties is reduced.

For the full sample, the median offsets remain approximately constant with distance from the pointing centre, with no systematic radial trend in either coordinate. The same behaviour is seen for the high-S/N subset, although the uncertainties increase in the outermost bins owing to the smaller number of sources. The persistence of the declination offset at high S/N indicates that it is not driven primarily by positional uncertainties of the faintest sources. These tests therefore provide no evidence for a field-dependent astrometric distortion across the region considered here.

\section{Catalogue description}
\label{app:catalogue}

The COSMOS 144\,MHz source catalogue is released in electronic format through CDS. It presents the basic properties fitted by {\sc PyBDSF} and used throughout this paper. Table~\ref{tab:catalogue_columns} lists the catalogue columns.

\begin{table*}[t]
\centering
\caption{Description of the principal columns in the released COSMOS 144\,MHz source catalogue. }
\label{tab:catalogue_columns}
\small
\begin{tabular}{llll}
\hline
Column name & Units & Description\\
\hline

Source\_ID & --  & The radio name of the source in the format "COSMOS144MHzJHHMMSS.s+DDMMSS" \\

RA & deg & Right ascension (J2000) \\

E\_RA & arcsec & Uncertainty in right ascension \\

DEC & deg & Declination (J2000) \\

E\_DEC & arcsec & Uncertainty in declination \\

Total\_flux & mJy & Integrated Stokes I flux density at 144 MHz \\

E\_Total\_flux & mJy & Uncertainty in integrated Stokes I flux density \\

Peak\_flux & mJy beam$^{-1}$ & Peak Stokes I flux density at 144 MHz \\

E\_Peak\_flux & mJy beam$^{-1}$ & Uncertainty in peak flux density \\

Maj & arcsec & Fitted FWHM of the major axis \\

E\_Maj & arcsec & Uncertainty in fitted major axis \\

Min & arcsec & Fitted FWHM of the minor axis \\

E\_Min & arcsec & Uncertainty in fitted minor axis \\

PA & deg & Position angle of the fitted major axis (east of north) \\

E\_PA & deg & Uncertainty in fitted position angle \\

DC\_Maj & arcsec & Deconvolved FWHM of the major axis \\

E\_DC\_Maj & arcsec & Uncertainty in the deconvolved major axis \\

DC\_Min & arcsec & Deconvolved FWHM of the minor axis \\

E\_DC\_Min & arcsec & Uncertainty in the deconvolved minor axis \\

DC\_PA & deg & Deconvolved position angle (east of north) \\

E\_DC\_PA & deg & Uncertainty in the deconvolved position angle \\

Isl\_Total\_flux & mJy & Integrated Stokes I flux density of the source island \\

E\_Isl\_Total\_flux & mJy & Uncertainty in the island integrated flux density \\

Isl\_rms & mJy beam$^{-1}$ & Average local rms noise of the source island \\

S\_Code & -- & {\sc PyBDSF}  source classification (S = single Gaussian, C = single Gaussian in a blended island, \\ 
& & M = multi-Gaussian source) \\

$ID_{\rm 3\,GHz}$ & -- &VLA-COSMOS 3\,GHz counterpart identifier \\ 

$S_{\rm 3\,GHz}$ & $\mu$Jy &VLA-COSMOS 3\,GHz flux density ($\mu$Jy) \\

$\alpha_{144\,\mathrm{MHz}}^{3\,\mathrm{GHz}}$ & -- & 
144\,MHz--3\,GHz spectral index, calculated using
$S_{\rm P}$ for unresolved sources and $S_{\rm T}$
for resolved sources, assuming $S_\nu\propto\nu^{-\alpha}$ \\

\hline
\end{tabular}
\end{table*}

\section{Radio surveys of the COSMOS field}
\label{app:radio_surveys}

\begin{figure*}[t]
    \centering
    \includegraphics[width=\textwidth]{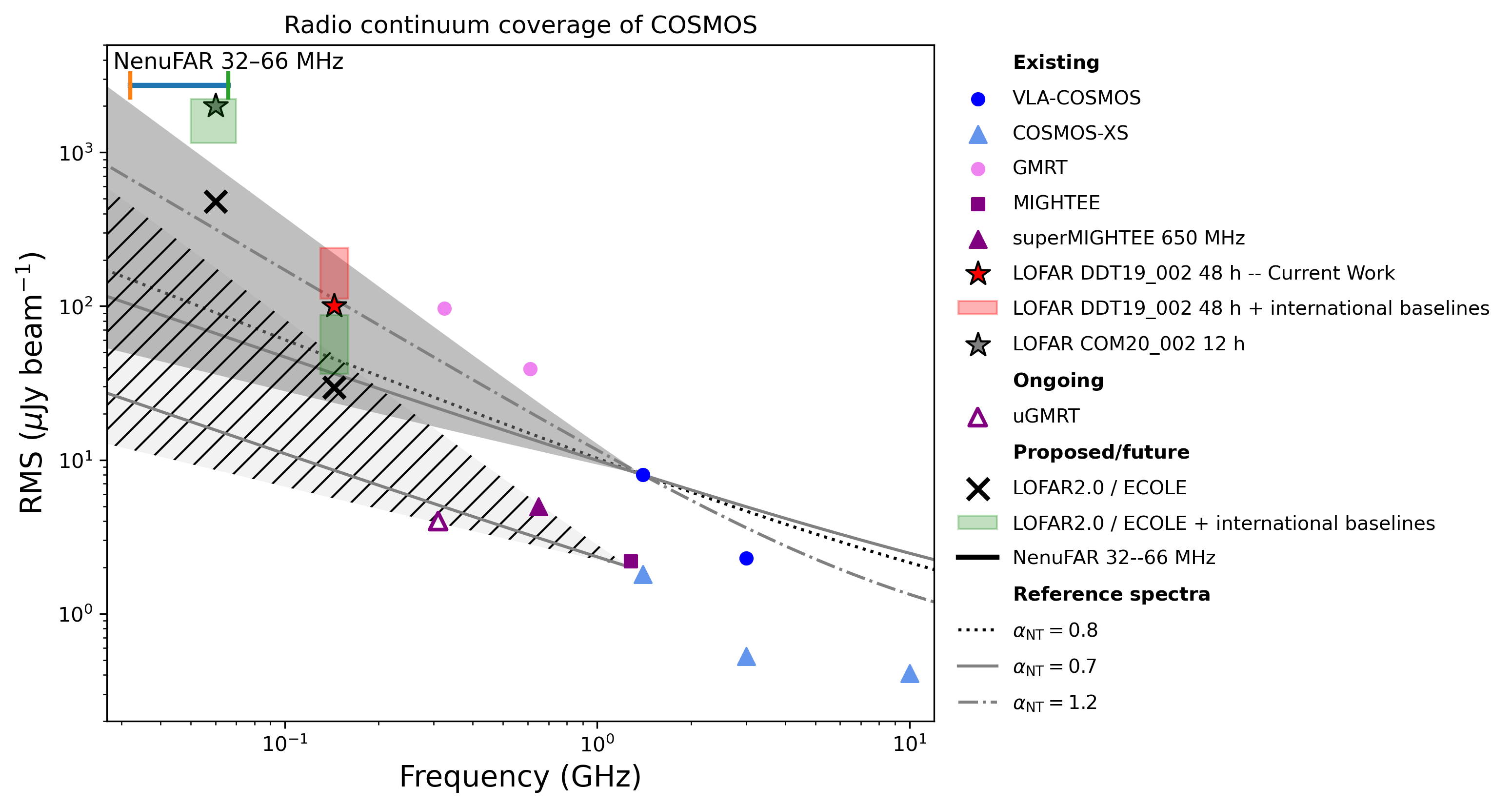}
  \caption{Comparison of the sensitivities of radio-continuum observations of the COSMOS field as a function of observing frequency. Existing observations are shown together with additional uGMRT coverage and the projected sensitivities of the proposed LOFAR2.0/ECOLE and NenuFAR observations. The existing surveys include the VLA-COSMOS surveys \citep{Schinnerer2007,Schinnerer2010,Smolcic2017a}, COSMOS-XS \citep{vanDerVlugt2021}, the GMRT COSMOS observations \citep{Tisanic2019}, MIGHTEE \citep{Heywood2022,Hale2025}, and the superMIGHTEE uGMRT observations at 650~MHz \citep{Lal2025}, together with the 144 MHz LOFAR observations presented in this work (red star) and the existing 60 MHz LOFAR observations. The grey lines show reference radio spectra with non-thermal spectral indices. The shaded grey and hatched regions show the range of two-component radio spectra with non-thermal spectral indices $\alpha_{\rm NT}=0.5$--$1.5$, assuming a 10\% thermal contribution ($\alpha_{\rm th}=0.1$). The grey region is normalised to $8\,\mu$Jy\,beam$^{-1}$ at 1.4 GHz, while the hatched region is normalised to $2\,\mu$Jy\,beam$^{-1}$ at 1.28 GHz.
  }
    \label{fig:future_surveys}
\end{figure*}

Figure~\ref{fig:future_surveys} places the 144\,MHz observations presented in this work in the context of existing radio-continuum observations of the COSMOS field and the projected sensitivity of future low-frequency observations.

\end{document}